\documentclass[aps, prb,reprint,10pt,superscriptaddress,amssymb,amsfonts]{revtex4-1}

\usepackage{amsmath,amssymb,amsthm,amscd,latexsym,epsfig,bm}
\usepackage[colorlinks,bookmarks=false,citecolor=blue,linkcolor=red,urlcolor=blue]{hyperref}
\usepackage{verbatim}
\usepackage{soul}

\def\be{\begin{equation}}
\def\ee{\end{equation}}

\newcommand{\zz}{\mathbb{Z}_2}

\def\cX{\mathcal{X}}
\def\cM{\mathcal{M}}

\def\ba{\boldsymbol{a}}
\def\fe{\mathfrak{e}}
\def\fm{\mathfrak{m}}

\begin{document}

\title{Fracton topological order via coupled layers}

\author{Han Ma}
\affiliation{Department of Physics, University of Colorado, Boulder, Colorado 80309, USA}
\affiliation{Center for Theory of Quantum Matter, University of Colorado, Boulder, Colorado 80309, USA}
\author{Ethan Lake}
\affiliation{Department of Physics and Astronomy, University of Utah, Salt Lake City, Utah 84112, USA}
\author{Xie Chen}
\affiliation{Department of Physics and Institute for Quantum Information and Matter, California Institute of Technology, Pasadena, California 91125, USA}
\author{Michael Hermele}
\affiliation{Department of Physics, University of Colorado, Boulder, Colorado 80309, USA}
\affiliation{Center for Theory of Quantum Matter, University of Colorado, Boulder, Colorado 80309, USA}

\date{\today}

\begin{abstract}

In this work, we develop a coupled layer construction of fracton topological orders in $d=3$ spatial dimensions. These topological phases have sub-extensive topological ground-state degeneracy and possess excitations whose movement is restricted in interesting ways. Our coupled layer approach is used to construct several different fracton topological phases, both from stacked layers of simple $d=2$ topological phases and from stacks of $d=3$ fracton topological phases. This perspective allows us to shed light on the physics of the X-cube model recently introduced by Vijay, Haah, and Fu, which we demonstrate can be obtained as the strong-coupling limit of a coupled three-dimensional stack of toric codes.  We also construct two new models of fracton topological order: a semionic generalization of the X-cube model, and a model obtained by coupling together four interpenetrating X-cube models, which we dub the ``Four Color Cube model.''
The couplings considered lead to fracton topological orders via mechanisms we dub ``p-string condensation'' and ``p-membrane condensation,'' in which strings or membranes built from particle excitations are driven to condense. This allows the fusion properties, braiding statistics, and ground-state degeneracy of the phases we construct to be easily studied in terms of more familiar degrees of freedom. Our work raises the possibility of studying fracton topological phases from within the framework of topological quantum field theory, which may be useful for obtaining a more complete understanding of such phases.  

\end{abstract}

\maketitle

\section{Introduction}

Quantum phases of matter in $d$ spatial dimensions are said to have topological order\cite{wen1989,wen1990,wen1990groundstate,wen2013} when they have an excitation gap, exhibit degenerate ground states on the torus (or other topologically nontrivial manifolds) that cannot be distinguished by local measurements, and support excitations that can be localized in space but cannot be created by a local process.\footnote{More precisely, we are describing non-invertible topological orders, a class that excludes symmetry-protected topological phases such as the $S=1$ Haldane chain and topological band insulators, as well as integer topological phases such as the integer quantum Hall liquids. We note that in some cases the topological excitations are not point objects, but can be localized to a $c$-dimensional subspace where $c<d$.}
In $d=2$, there is by now a good understanding of topological order, ranging from its realization in fractional quantum Hall liquids\cite{tsui82,laughlin83} and in bosonic models,\cite{chakraborty89,read91,wen91,sachdev92,balents99,senthil00,moessner01a,moessner01b,balents02,kitaev03,levin05} to topological quantum field theories such as Chern-Simons theory,\cite{witten1989quantum} to the general framework of modular tensor category theory that describes topological orders in bosonic systems.\cite{kitaev2006anyons}

It has recently become apparent that exotic new types of topological order exist in three spatial dimensions.\cite{chamon2005quantum, bravyi2011topological, haah2011local, bravyi2011energy, yoshida2013exotic,bravyi2013quantum,haah2013commuting,haah2014bifurcation, vijay2015newkind, vijay2016fracton,pretko2016subdimensional,pretko2016generalized} A number of exactly solvable quantum spin models have been found with long-range entangled ground states, a gap to local excitations, and excitations carrying non-trivial topological charge that cannot be created locally.  These properties are shared with familiar two-dimensional topological orders and with three-dimensional discrete gauge theories, including twisted (Dijkgraaf-Witten) gauge theories.\cite{dijkgraaf1990topological}  Unlike those examples, however, these exotic three dimensional states have point-like excitations that are confined to move in zero, one or two-dimensional subspaces. We refer to such excitations as zero-, one- and two-dimensional particles, respectively.

The zero-dimensional particles are dubbed fractons,\cite{vijay2015newkind} and are fundamentally immobile in the sense that they cannot be created at ends of one-dimensional string operators. Instead, in some models fractons are created at corners of two-dimensional membrane operators,\cite{chamon2005quantum,vijay2015newkind,vijay2016fracton} while in other models they are created at corners of fractal operators.\cite{haah2011local,yoshida2013exotic} In either case, a process destroying a single isolated fracton must create more than one fracton elsewhere in space, so that individual fractons cannot simply move from one point to another on their own. 

Many basic questions about fracton topological orders remain open. One such question is whether some fracton topological orders can be related to and understood in terms of more familiar quantum phases of matter and their degrees of freedom. For example, the existence of two-dimensional particles could originate from a weakly coupled stack of $d=2$ topologically ordered layers. Along the same lines, composites of excitations in two intersecting layers can be confined to move in one dimension, along the intersection line of the layers, and composites of excitations in three intersecting layers are completely immobile. Not all the features of fracton topological orders can be explained by simply stacking two dimensional topological orders. However, since some properties are similar to those of simple stacks, it is natural to ask whether we can take decoupled $d=2$ topologically ordered layers, and couple them so as to obtain $d=3$ fracton topological orders.

In this paper, we show that some fracton topological orders can be understood by suitably coupling layers of familiar $d=2$ topologically ordered systems.  The coupling can be understood as a condensation of one-dimensional extended objects formed from particle excitations of the $d=2$ layers, which we dub particle strings or ``p-strings.'' This provides a simple understanding of the properties of excitations, and ground state degeneracy, of the resulting fracton state.  We also take this idea one step further, and couple together $d=3$ fracton topological orders to obtain new $d=3$ fracton topological orders by condensing two-dimensional membranes built from point particles, dubbed p-membranes.

This paper is organized as follows. In Sec. \ref{SC}, we show how to obtain a certain type of fracton topological order, realized in the X-cube model of Ref.~\onlinecite{vijay2016fracton}, by coupling together layers of toric codes\cite{kitaev03} covering the simple cubic lattice. The coupling is a $ZZ$ exchange interaction, and in the strong coupling limit, we reproduce the Hamiltonian of the X-cube model at sixth order in perturbation theory. We also consider $XX$ coupling, where we obtain the usual $d=3$ toric code model in the strong coupling limit.

The strong coupling analysis suggests that the X-cube topological order can be understood in terms of the degrees of freedom of $d=2$ toric code layers. Section~\ref{MCa} examines this relationship through the lens of p-string condensation at intermediate coupling strengths. The X-cube model supports one-dimensional ``electric'' particle excitations and zero-dimensional ``magnetic'' fractons.

We show that the one-dimensional particles are formed from pairs of toric code $e$-particles, and that fractons are created at the ends of open p-strings. $m$-particles survive in the X-cube model as bound states of two fractons. The resulting perspective also allows us to easily calculate the ground state degeneracy (GSD) of the X-cube model, which we do in Sec.~\ref{sec:gsd}. In Sec.~\ref{MCc}, we apply a similar intermediate coupling picture to the case of $d=2$ toric codes coupled by $XX$ coupling, where condensation of composites of two $e$-particles leads to a standard $d=3$ toric code phase.

In Sec.~\ref{CXC}, we introduce a new type of fracton topological order obtained via p-membrane condensation in a system of four interpenetrating X-cube models. We dub the resulting exactly sovable model the ``Four Color Cube (FCC)'' model.  The FCC model has an electric-magnetic self-duality, and all its excitations can be obtained as composites of immobile fractons. We describe the properties of the FCC model's excitations, and calculate its GSD on the three-torus, which we find to be $\log_2GSD_{FCC} = 32L-24$.

In Sec.~\ref{SXC} we construct a semionic version of the X-cube model, obtained by coupling together stacks of models with doubled semion topological order. The main new feature in the semionic version of the X-cube model is the addition of nontrivial ``braiding statistics'' between the one-dimensional excitations. We expect that our method of obtaining the semionic X-cube model can be readily extended to construct fracton phases from coupled stacks of more general types of topological order. 

The paper concludes in Sec.~\ref{disc}, where we discuss future directions and open questions raised by our work. Technical details are given in two appendices. 

\section{X-cube model from toric code layers: strong coupling}
\label{SC} 

Our starting point is a model of coupled layers of $d=2$ toric codes. In this section, we consider the limit of strong coupling, and show that our model reduces exactly to the X-cube Hamiltonian in this limit, using standard techniques of degenerate perturbation theory. We also show that, in the same model but with a different form of coupling, upon taking the strong coupling limit we obtain the conventional $d=3$ toric code model.

We begin by describing the system in the decoupled limit. We consider \emph{three} independent stacks of square-lattice toric codes along the $\mu = x,y,z$ directions of the cubic lattice. The toric codes cover the cubic lattice in such a way that two Ising spins reside on every cubic link. For example, a link oriented in the $x$-direction is contained in one $xy$ plane and one $xz$ plane. One of the spins on the link is part of an $xy$-plane toric code, while the other participates in a $xz$-plane toric code.

Before describing how to couple the toric codes, we establish some notation.  We label cubic lattice vertices by $i,j$, links by $\ell$, square plaquettes by $p$, and $\{100\}$, $\{010\}$ and $\{001\}$ lattice planes by $P$. Lattice directions are indicated by $\mu = x,y,z$, we write $\ell = (i, \mu)$ for the link extending from $i$ in the $+\mu$ direction, and we refer to links as $\mu$-links when we want to indicate their direction.  Sometimes it is convenient to indicate links by nearest-neighbor pairs of sites $ij$.  Each plaquette $p$ has an orientation denoted $o(p) = x,y,z$, which specifies the direction normal to $p$. Similarly, the orientation of the plane $P$ is written $o(P)$.

On each link we place two Ising spins.  $Z$ and $X$ Pauli operators for the spins on the link $\ell = (i, \mu)$ are written $Z^{\nu}_{(i, \mu)}, X^{\nu}_{(i, \mu)}$.  The superscript $\nu = x,y,z$ ($\nu \neq \mu$) indicates the orientation of the toric code plane in which the spin participates.  That is, each spin is a member of a $d=2$ toric code on plane $P$ that contains $\ell$, and with $\nu = o(P)$.

The toric code Hamiltonian for plane $P$ is written
\begin{equation}
H^{TC}_{P} = - \sum_{i \in P} A^{o(P)}_i - \sum_{p \in P} B_p \text{,}
\label{TC_sq}
\end{equation}
where we introduced the usual vertex operators
\begin{equation}
A^{\mu}_i  = \prod_{ij \perp \mu} Z^{\mu}_{i j} \text{.}
\end{equation}
Here, the product is over the four links touching $i$ and perpendicular to the direction $\mu$. We also introduced plaquette operators
\begin{equation}
B_p = \prod_{\ell \in p} X^{o(p)}_{\ell} \text{.}
\end{equation}
We note that $B_p$ does not carry a superscript indicating its orientation, as this is already implicit upon specifying $p$.  

Now we couple together the toric code layers with the Hamiltonian
\begin{equation}
H = \sum_P H^{TC}_P - J_z \sum_{\ell} Z^{\mu_1}_{\ell}  Z^{\mu_2}_{\ell} \text{.}
\end{equation}
Here, $\mu_1$ and $\mu_2$ are the two lattice directions perpendicular to $\ell$.  The two spins on each link now interact via a $Z Z$ exchange interaction with coefficient $J_z > 0$.  When $J_z = 0$ we have decoupled toric code layers.

We now take the limit $J_z \to \infty$ and treat $H^{TC} = \sum_P H^{TC}_P$ as a perturbation.  First ignoring the perturbation, we have an extensively degenerate ground state space, where any spin configuration in the $Z$ basis satisfying the constraint $Z^{\mu_1}_{\ell} = Z^{\mu_2}_{\ell}$ is a ground state.  The ground space on link $\ell$ is that of an Ising spin with Pauli operators
\begin{eqnarray}
{\cal Z}_{\ell} &\equiv& Z^{\mu_1}_{\ell} = Z^{\mu_2}_{\ell} \\
{\cal X}_{\ell} &\equiv& X^{\mu_1}_{\ell} X^{\mu_2}_{\ell} \text{,}
\end{eqnarray}
which commute with the $J_z$ coupling term.  Any operator acting within the many-body ground space can be written in terms of these operators.

\begin{figure}
\includegraphics[width=.4\textwidth]{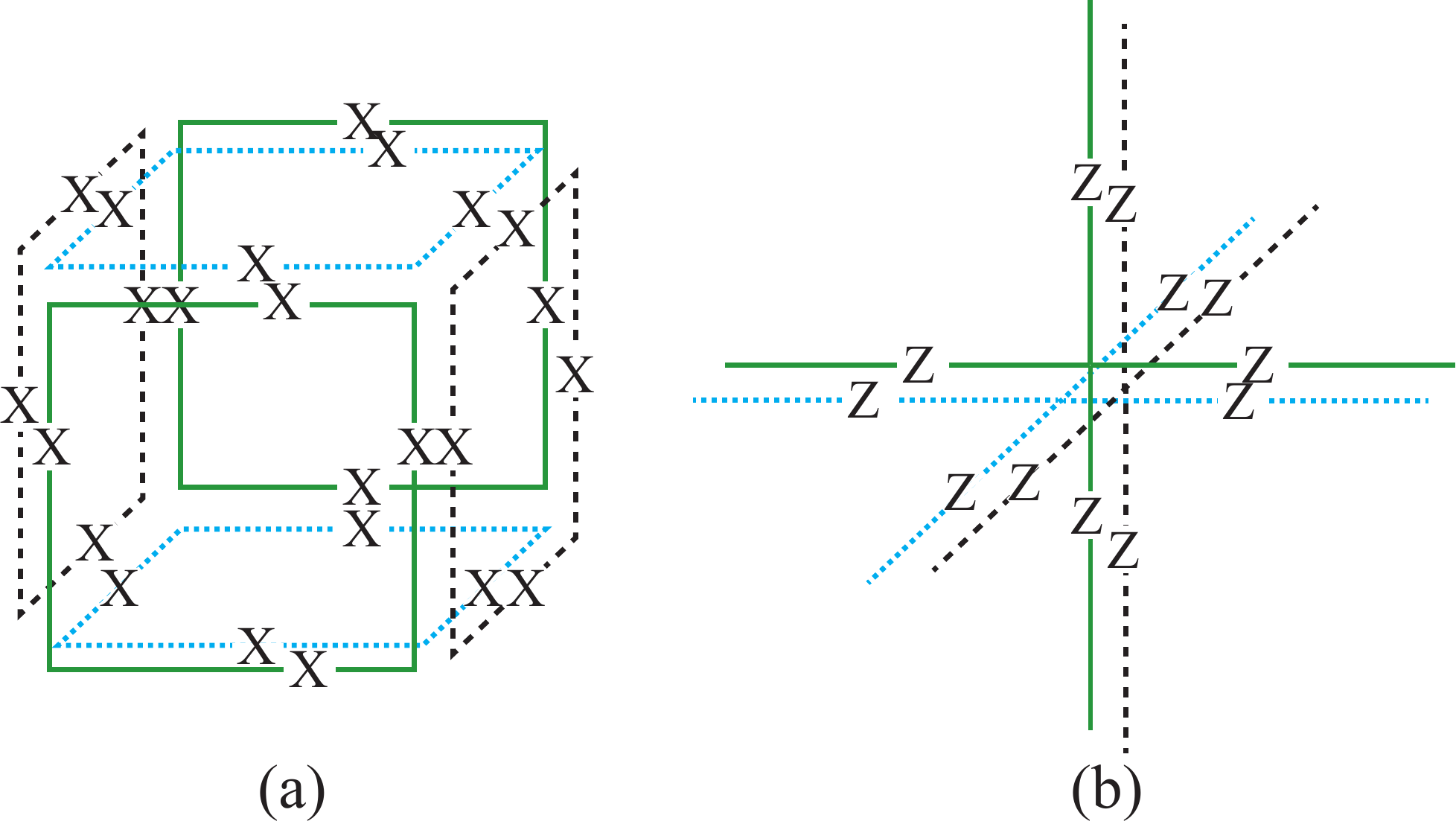}
\caption{Illustration of cube and vertex terms in the X-cube model related to plaquette and vertex terms from toric code layers. \label{fig:terms}}
\end{figure}

The ground state space is split by an effective Hamiltonian we obtain using Brillouin-Wigner degenerate perturbation theory.  The details are described in Appendix~\ref{app:perturbation-theory}; it is necessary to take care of some technical issues in order to go beyond leading order. 
We obtain the X-cube Hamiltonian at sixth order in perturbation theory:
\begin{equation}
H_{XC} = - \sum_{i} \sum_{\mu = x,y,z} {\cal A}^{\mu}_i - K \sum_c {\cal B}_c \text{.} \label{eqn:xcube}
\end{equation}
where the relation between terms in the X-cube model and toric code layers is shown in Fig. (\ref{fig:terms}). Here, we have dropped constant terms, and we defined
\begin{equation}
{\cal A}^{\mu}_{i} = \prod_{ij \perp \mu} {\cal Z}_{i j} \text{.}
\end{equation}
Moreover, $c$ labels elementary cubes of the lattice, and
\begin{equation}
{\cal B}_c = \prod_{\ell \in c} {\cal X}_{\ell} \text{,}
\end{equation}
where the product is over the 12 edges of the cube $c$. The coupling of the cube term is $K = C / J_z^5$, where $C > 0$ is a constant factor that can be computed if desired following the discussion of Appendix~\ref{app:perturbation-theory}.  We have not computed $C$ because its value plays no role in our discussion.

To summarize the perturbation theory analysis, the vertex term of $H^{TC}$ contributes at first order and gives the vertex term (first term) of the X-cube Hamiltonian.  This is simple to understand, as projection to the ground state space simply replaces $Z^{\mu}_{\ell}$ with ${\cal Z}_{\ell}$.  The cube term is a product of six $B_p$'s, where each $B_p$ operator anticommutes with the $J_z$ term on the four links in the perimeter of $p$.  To obtain a non-trivial operator within the ground state space as a product of $B_p$'s, we have to take a product over plaquettes forming a closed surface.  The smallest such surface is a single cube with six faces, so the cube term is the lowest-order such contribution arising in perturbation theory.  This is described in more detail in Appendix~\ref{app:perturbation-theory}, where it is also shown that no other terms arise between first and sixth order.

We have shown that $H$ interpolates between decoupled $d=2$ toric codes when $J_z = 0$, and the X-cube Hamiltonian when $J_z \to \infty$. This suggests that the topological order of the X-cube model can be understood in terms of degrees of freedom of the decoupled toric codes. To develop this idea, we need to consider the effect of the $J_z$ term at weak and intermediate coupling, which is done in Sec.~\ref{MC}.

We conclude this section with a brief discussion of the effect of replacing the $ZZ$ coupling with $XX$ coupling, specifically adopting the Hamiltonian 
\begin{equation}
H_{XX} = \sum_P H^{TC}_P - J_x \sum_{\ell} X^{\mu_1}_{\ell} X^{\mu_2}_{\ell} \text{.}
\end{equation}
Again considering the $J_x \to +\infty$ limit, the single-site ground space is that of an Ising spin with Pauli operators ${\cal X}_{\ell} \equiv X^{\mu_1}_{\ell} = X^{\mu_2}_{\ell}$ and ${\cal Z}_{\ell} = Z^{\mu_1}_{\ell} Z^{\mu_2}_{\ell}$. Degenerate perturbation theory results in the usual $d=3$ toric code model,
\begin{equation}
H_{3dTC} = - \sum_p {\cal B}_p - \tilde{K} \sum_{i} {\cal A}_i \text{,}
\end{equation}
where ${\cal B}_p = \prod_{\ell \in p} {\cal X}_{\ell}$, and ${\cal A}_i = \prod_{j} {\cal Z}_{ij}$, with the latter product over the six links touching $i$. The coefficient of the vertex term satisfies $\tilde{K} \propto J_x^{-2}$. The fact that we obtain the $d=3$ toric code model when $J_x \to \infty$ can be understood coming from the limit of weak $J_x$ as a condensation of bound states $e_1 e_2$, where $e_1, e_2$ are $e$-particles in two intersecting toric code layers. This is described in Sec.~\ref{MCc}.

\section{X-cube model from toric code layers: intermediate coupling}
\label{MC}

While it is suggestive, the fact that we obtain the X-cube model upon strongly coupling toric code layers does not directly relate the properties of the X-cube model to the properties of decoupled toric codes.  Motivated by the strong coupling analysis, here we consider small and intermediate values of $J_z$, and show that the topological order of the X-cube model can be obtained from decoupled toric codes by condensation of extended one-dimensional objects we dub ``p-strings.''  This allows us to describe properties of the X-cube model in terms of the degrees of freedom of decoupled toric code layers.  In particular, we use this condensation picture to recover the properties of the X-cube model's topological excitations, and to give a simple computation of the ground state degeneracy on a three-torus. In Sec.~\ref{MCc}, we consider a similar picture for toric code layers with $XX$ coupling, where condensation of bound pairs of two $e$ particles leads to a description of $d=3$ toric code topological order.

\subsection{Fracton topological order from p-string condensation}
\label{MCa}

\begin{figure}
\includegraphics[scale=.8]{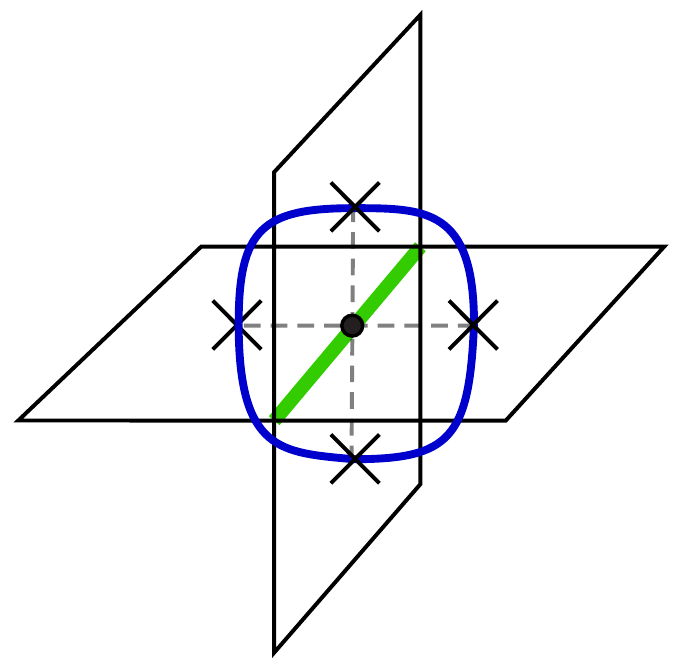}
\caption{\label{mini_ploop} (Color online) An elementary p-string which forms the building block of the p-string condensate. The green thick link denotes an action of $Z_\ell^xZ_\ell^y$, which creates four $m$ particles (shown as black x's connected by the  dashed lines). The black dot shows the location of the physical spin ${\cal Z}_\ell$($=Z_\ell^x=Z_\ell^y$). The blue string represents the p-string, which connects the $m$ particles on the perimeter of the membrane. }
\end{figure}

\begin{figure}
\includegraphics[scale=.8]{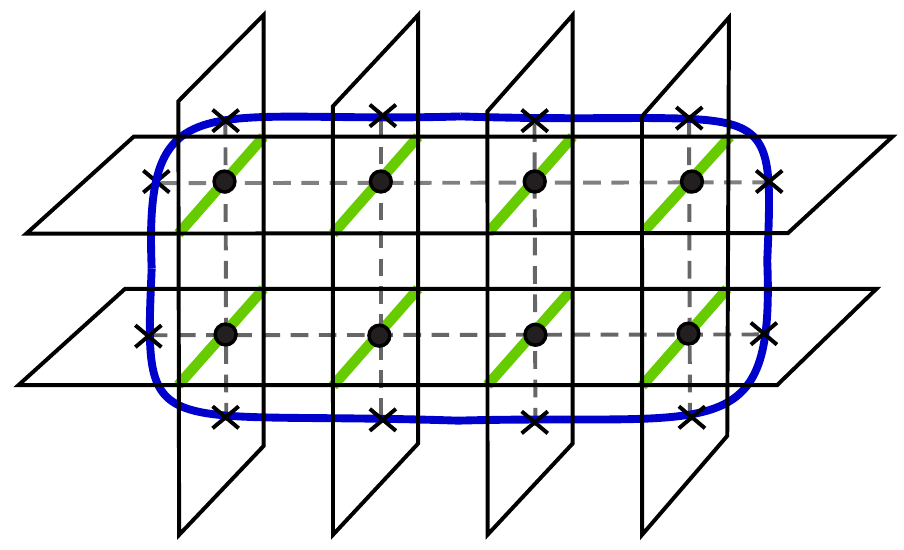}
\caption{\label{big_ploop} (Color online) A larger p-string, obtained by acting with $Z_\ell^xZ_\ell^z$ operators along the links orthogonal to a rectangular membrane (marked in green).}
\end{figure}

Since the $J_z$ coupling term does not commute with the $B_p$ term in the original toric code Hamiltonian, acting with it creates toric code $m$ particles, which occur on plaquettes that violate the $B_p$ terms in $H_P^{TC}$. In particular, acting with the coupling operator $Z_\ell^{\mu_1}Z_\ell^{\mu_2}$ creates two pairs of $m$ particles on the four plaquettes touching the link $\ell$ (Fig~\ref{mini_ploop}).

If we represent each $m$ particle by a line segment normal to the plane in which it moves, the line segments for the four $m$ particles created by $Z_\ell^{\mu_1}Z_\ell^{\mu_2}$ can be joined into a closed string. We refer to this string as a p-string, where ``p'' stands for particle, as it is built out toric code particle excitations. An elementary p-string created by the application of a single $Z_\ell^{\mu_1}Z_\ell^{\mu_2}$ operator is shown in Fig. \ref{mini_ploop}. Acting with a collection of $Z_\ell^{\mu_1}Z_\ell^{\mu_2}$ operators on links orthogonal to a rectangular membrane creates larger p-strings, as shown in Fig. \ref{big_ploop}. 

As we increase $J_z$ from zero, at some point we expect to induce a condensation of the p-strings created by the $J_z$ exchange interaction. Upon condensation, p-strings appear at all sizes and propagate freely through the system, driving a confinement transition in the electric sectors of the toric code layers. In particular, we will see that this condensation process leads to the confinement of individual $e$ particles, while bound pairs of $e$ particles on intersecting planes survive the condensation process. 

\begin{figure}
\includegraphics{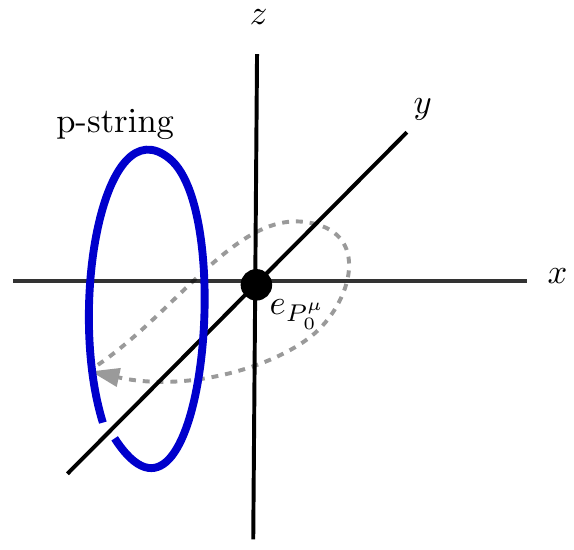}
\caption{\label{braiding_picture} (Color online) A braiding process between a p-string in the condensate and an $e_{P_0^\mu}$ particle. The process is drawn in a ``continuum limit,'' where we do not show the individual $m$ particles making up the p-string. During the braiding process the right side of the p-string is held fixed, while the left side sweeps out the motion indicated by the gray arrow.}
\end{figure}

In order to examine what happens to the toric code $e$ particles under the p-string condensation, we will examine how $e_P$ particles braid with the p-string condensate, where $e_{P}$ denotes a plane-$P$ toric code $e$ particle (a violation of the $A_i^{o(P)}$ term in $H^{TC}_{P}$). 
The kind of braiding process we need to look at is one in which a p-string winds around a single $e_{P_0^\mu}$ particle in a particular plane $P_0^{\mu}$ with $o(P_0^\mu)=\mu$, as illustrated in Fig. \ref{braiding_picture}.

During the braiding process, we move the p-string around the $e_{P_0^\mu}$ excitation by keeping the right part of the loop fixed and moving the left part around the $e_{P_0^\mu}$ excitation as shown by the path of the gray arrow in Figure~\ref{braiding_picture}. Since p-strings are composed of $m$ particles, each intersection of the loop with a plane $P$ defines the location of an $m_P$ particle in that plane. We set the location of the $e_{P_0^\mu}$ excitation as the origin of our coordinate system, and label the three planes containing the origin by $P^{\nu}_0$ for $\nu = x,y,z$, and $o(P^\nu_0) = \nu$. In particular, the $xy$ plane containing the origin is $P_0^z$. Tracking the intersection of the p-string with $P_0^z$, we see that moving the p-string induces a braiding of an $m_{P_0^z}$ particle with  $e_{P_0^\mu}$, which contributes a phase of $\theta_{e_{P_0^\mu},m_{P_0^z}} = \pi \delta_{\mu,z}$ to the braiding of the p-string with $e_{P_0^\mu}$. Intersections of the p-string with other $xy$ planes $P$ not containing the origin ($o(P) = z$ but $P \neq P_0^z$) do not contribute, because in that case $\theta_{e_{P_0^z},m_P}=0$.

Similar arguments apply to for $P^x_0$ and $P^y_0$, whose intersections with the p-string are $m_{P_0^x}$ and $m_{P_0^y}$ particles. These particles are braided around $e_{P_0^\mu}$ during the p-string braiding process, contributing phases of $\theta_{e_{P_0^\mu},m_{P_0^x}} = \pi \delta_{\mu,x}$ and $\theta_{e_{P_0^\mu},m_{P_0^y}} = \pi \delta_{\mu,y}$, respectively. 
Putting everything together, we see that 
\begin{equation}
\theta_{e_{P_0^\mu},\text{p-string}} = \sum_{\nu = x,y,z} \theta_{e_{P_0^\mu},m_{P_0^\nu}} = \pi \text{.}
\end{equation}
Because this braiding phase is non-trivial, all individual $e_P$ particles become confined after inducing the p-string condensation. 

However, the condensation process does not completely confine the electric sector excitations of the original decoupled $J_z =0$ theory. Instead, it allows bound-state pairs of $e$ particles on intersecting planes to remain deconfined. Indeed, consider the bound state of two distinct $e_P$ excitations $e_{P_i}e_{Q_i}$, with $P\neq Q$, $o(P) \neq o(Q)$ so that the planes intersect, and where $e_{P_i}$ denotes an $e$ excitation located on vertex $i$ in plane $P$. These bound states have trivial braiding with the p-string condensate, so these composites of two $e$ particles are deconfined even in the $J_z \rightarrow \infty$ limit\footnote{Pairs like $e_Pe_Q$ with $o(P)=o(Q)$ are also confined, since p-strings can pass between the two $e_P$ excitations, giving a $\pi$ braiding phase that confines them.}. We denote these bound-state excitations as
\be \fe^i_\mu = e_{P_i^\nu}e_{P_i^\lambda},\ee
where $\mu,\nu,\lambda$ are three distinct directions, and $P_i^\mu$ denotes the plane containing the site $i$ and normal to the $\mu$ direction. We will often drop the $i$ superscript in $\fe^i_\mu$ when it will cause no confusion. 

An $\fe^i_\mu$ particle is able to move only along the $\mu$ direction, and so is a fundamentally one-dimensional particle. The $\fe_\mu$ excitations are precisely the 1d particles of Ref.~\onlinecite{vijay2016fracton}. We can use the fusion rule $e_P\times e_P = 1$ of the toric code to derive the fusion rules of the $\fe_\mu$ fractons:
\be
\fe^i_\mu \times \fe^i_\nu = \left\{ \begin{array}{ll} \fe^i_{o(\mu,\nu)}, & \mu \neq \nu \\
1, & \mu = \nu \end{array}\right. \text{,}
\ee
where $o(\mu,\nu)$ is the direction normal to both $\hat\mu$ and $\hat\nu$.

\begin{figure}
\includegraphics[scale=.9]{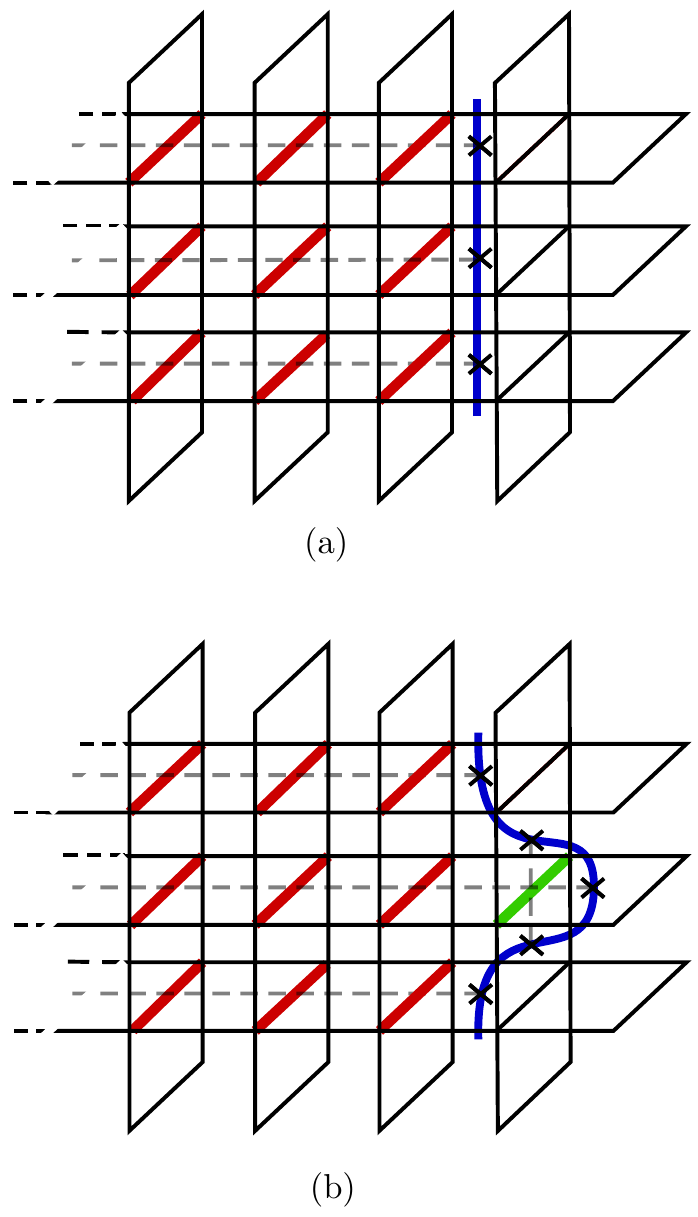}
\caption{\label{fig:finite_pstrings}(Color online) (a) An open p-string (dark blue), created at the edge of a series of $xy$ plane $m$-string operators (which act on the red links) stacked in the $z$-direction. The ends of the $m$-string operators are $m$-particles, and are marked with black crosses. (b) Acting with $Z_\ell^x Z_\ell^z$ on the green link creates a short $m$-string in the $yz$ plane and deforms the p-string.}
\end{figure}

Having discussed the electric excitations of the X-cube model, we now turn to an analysis of the magnetic excitations, which arise at ends of open p-strings. Working first in the decoupled limit, we consider the ``stack'' of $m_P$ particles in $xy$ planes shown in Fig.~\ref{fig:finite_pstrings}a. The following discussion holds more generally, but we focus on the particular geometry of Fig.~\ref{fig:finite_pstrings} to simplify the notation. Each $m_P$ particle is created at the end of an $m$-string operator
\begin{equation}
S_m(\gamma) = \prod_{\ell | \ell \cap \gamma \neq 0} Z^{z}_{\ell} \text{,}
\end{equation}
where $\gamma$ is a path lying in a $xy$ plane that intersects links transversely, and the product is over all links $\ell$ intersecting $\gamma$.

The stack of $m_P$ particles is taken to have finite extent in the $z$-direction, and can be represented as an open p-string, as shown in Fig.~\ref{fig:finite_pstrings} (a). Acting on the string with the coupling term $Z_\ell^{\mu_1}Z_\ell^{\mu_2}$ moves it around (Fig.~\ref{fig:finite_pstrings} (b)), 
but keeps the locations of the p-string endpoint fixed. This is because acting with $Z_\ell^{\mu_1}Z_\ell^{\mu_2}$ creates loops of p-string and therefore acting with $Z_\ell^{\mu_1}Z_\ell^{\mu_2}$ cannot change the $\zz$ flux of p-strings through any given cube.
Therefore, upon p-string condensation, the fixed endpoints of p-strings remain as excitations, but the one-dimensional ``bulks'' of the strings become tensionless, 
allowing them to fluctuate at all length scales without incurring any energetic penalty. This results in a condensate of p-strings, and the one-dimensional ``bulks'' of the strings cease to be physically observable objects.
The endpoints of the p-strings reside in cubes of the simple cubic lattice, and are the fracton excitations of the X-cube model. We will denote a fracton excitation supported at the cube $c$ by $\fm_c$. From the $\zz$ fusion rule of $m$ particles in the toric code (\emph{i.e.} the fusion rule $m_P\times m_P =1$), we see that the $\fm_c$ fractons also obey $\zz$ fusion rules. 

\begin{figure}
\includegraphics[scale=.75]{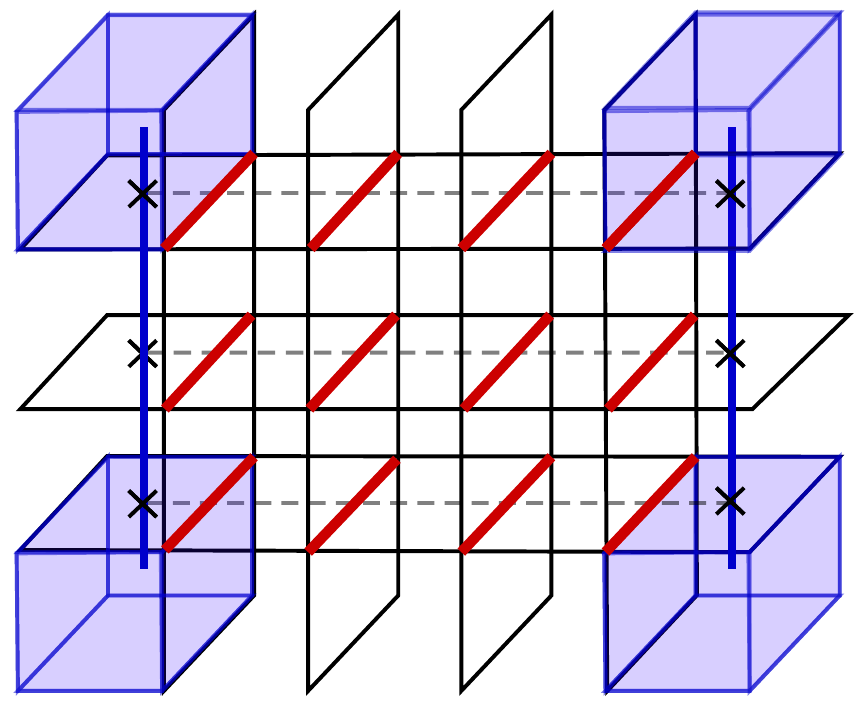}
\caption{ \label{fig:cube_membrane}(Color online) A membrane operator ${\cal M}_M$ which creates a pair of open p-strings harboring $\fm$ excitations (purple cubes) at their endpoints, formed by creating a vertical stack of $m$-strings. The red links possess $\cX_\ell$ eigenvalues of $-1$, and the dashed lines represent $m$-strings. }
\end{figure}

Figure~\ref{fig:cube_membrane} shows two stacks of $m_P$ particles, created by the operator
\begin{equation}
M_{\sigma} = \prod_{\ell | \ell \cap \sigma \neq 0} Z^z_{\ell} \text{,}
\end{equation}
where $\sigma$ is a rectangular membrane, and the product is over links $\ell$ cutting $\sigma$ transversely (drawn in red in Figure~\ref{fig:cube_membrane}). This operator can be viewed as a stack of $m$-strings, $M_{\sigma} = \prod_{\gamma \in \sigma} S_m(\gamma)$, with each string creating two $m_P$ particles on opposite sides of the membrane. Upon condensing p-strings, the operator $M_{\sigma}$ creates four fracton excitations at its corners. Indeed, in the strong coupling limit we can replace $Z^{\mu_1(\mu_2)}_{\ell}$ by ${\cal Z}_{\ell} \equiv Z^{\mu_1}_{\ell} = Z^{\mu_2}_{\ell} $ and
\begin{equation}
M_{\sigma} \to {\cal M}_{\sigma} = \prod_{\ell | \ell \cap \sigma \neq 0} {\cal Z}_{\ell} \text{,}
\end{equation}
which is simply a membrane operator of the X-cube model that creates fractons at its corners, as discussed in Ref.~\onlinecite{vijay2016fracton}.

\begin{figure}
\includegraphics[scale=.75]{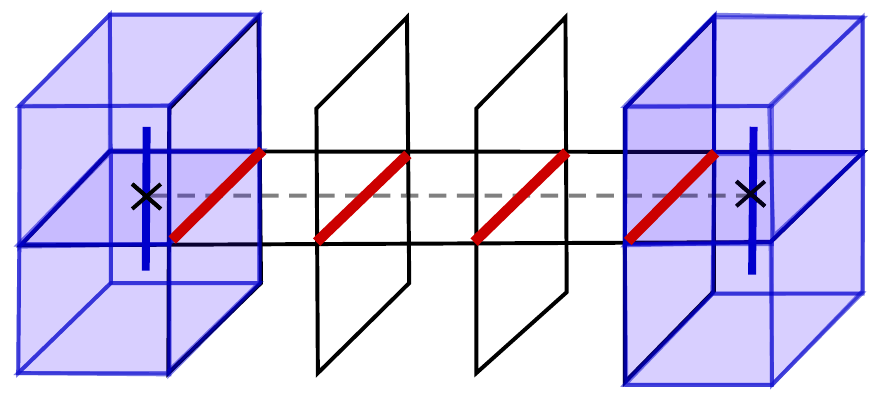}
\caption{\label{fig:cube_halfm}(Color online) A single $m$-string (dashed line) creates two $\fm_c$ excitations (purple cubes) at each of its ends, illustrating the fact that $m$-particles (black crosses) survive in the X-cube topological order as bound states of two $\fm_c$ fractons.}
\end{figure}

\begin{figure}
\includegraphics[scale=.75]{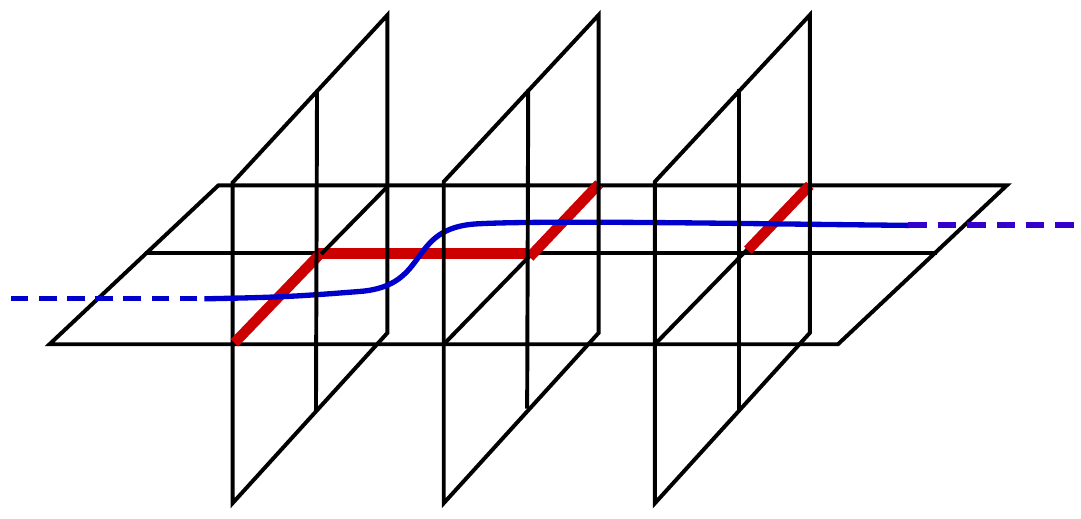}
\caption{\label{cube_string}(Color online) An illustration of the operator ${\cal S}_m(\gamma)$ for a section of a generic path $\gamma$. The red links denote links whose $\cX_\ell$ eigenvalue is $-1$, and the blue string denotes a section of the path $\gamma$.}
\end{figure}

It is instructive to consider the case where $\sigma$ contains only a single $m$-string, so that the membrane operator in the decoupled limit creates a pair of $m_P$ particles in a single layer, and $M_{\sigma} = S_m(\gamma)$. Taking the strong coupling limit, this allows us to define an $m$-string operator in the X-cube model by
\begin{equation}
{\cal S}_m(\gamma) = \prod_{\{\ell | \ell \cap \gamma \neq 0\}} {\cal Z}_\ell \text{.}
\end{equation}
In general, $\gamma$ is some path lying in a single $\{100\}$ plane that cuts links transversely, as shown in Fig.~\ref{cube_string}. In the X-cube model, the $m$-string ${\cal S}_m(\gamma)$ creates two $\fm_c$ fracton excitations at each end as shown in Fig.~\ref{fig:cube_halfm}. Therefore, $m$-particles survive in the X-cube topological order as bound states of two $\fm_c$ fractons.

In summary, our procedure of obtaining the X-cube model showcases two different ways of restricting the movement (or reducing the ``dimensionality'') of particles in topological phases. One mechanism, which occurs in the electric sector, is to bind together two particles which are free to move in two different (but intersecting) planes, resulting in a composite excitation free to move only in one dimension. The other mechanism, which occurs in the magnetic sector, is to {\it fractionalize} particles by ``breaking them apart'' into pairs of immobile particles. 

Our perspective on the X-cube model allows for a simple understanding of the statistical properties of $\fe$ and $\fm$ excitations. Without trying to give a general definition of statistical processes of particles moving in restricted dimensionalities, we note that \emph{any} statistical process involving only $\fe$ excitations must be trivial, because these excitations originate from a collection of toric code $e$ particles, which have trivial self and mutual statistics. The same statement holds for any statistical process involving only $\fm$ particles.

\begin{figure}
    \centering
    \includegraphics[width=.45\textwidth]{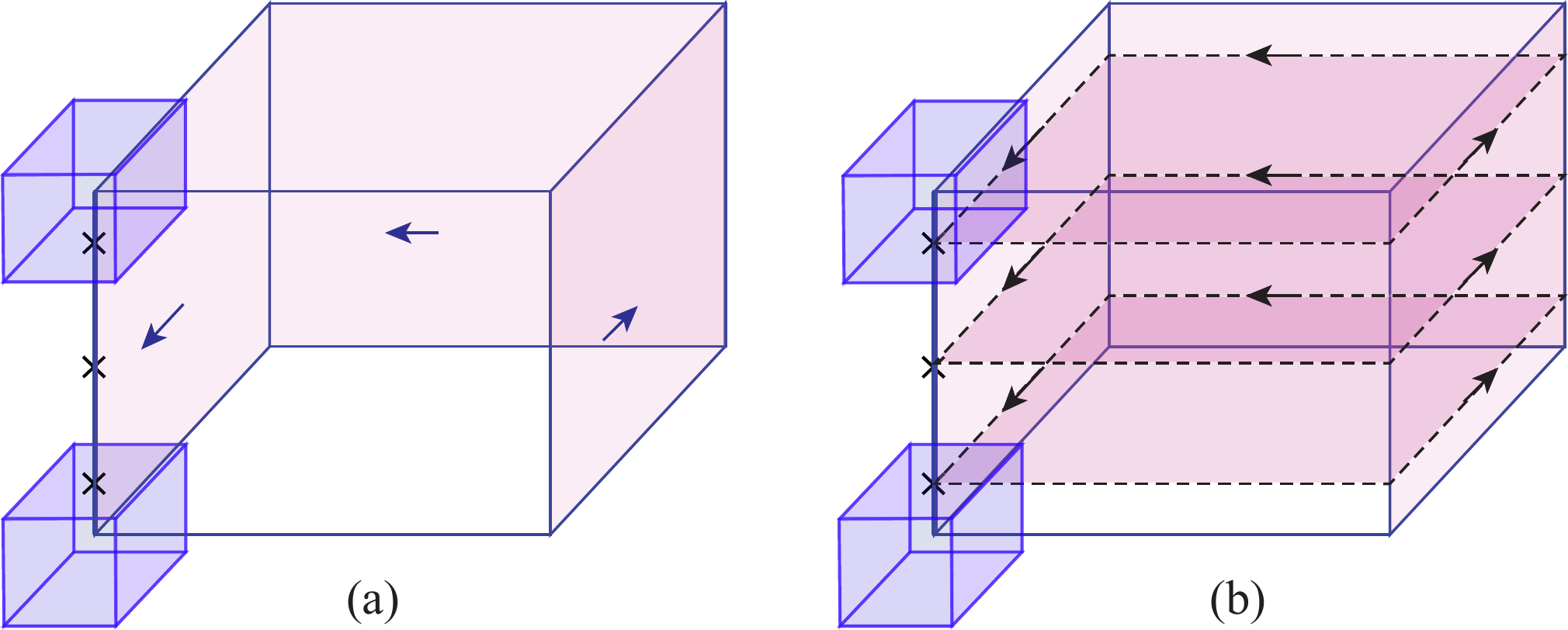}
    \caption{(Color online)(a) Two fractons are transported around a rectangular prism by acting with membrane operators (shaded areas) on the faces. (b) The membrane operator in (a) can be regarded as a stack of $m$ string operators (dashed lines), which transport a stack of $m$ particles (black crosses) around the prism.}
    \label{fig:braiding_x_cube}
\end{figure}

However, there are non-trivial statistical processes involving both $\fe$ and $\fm$ excitations.  Consider a rectangular prism, and let $\sigma_1, \dots, \sigma_4$ be four of its faces, excluding the two faces lying in $xy$ planes. Then the membrane operator
\begin{equation}
{\cal M}_{{\rm prism}} = {\cal M}_{\sigma_1} \cdots {\cal M}_{\sigma_4} 
\end{equation}
does not create any excitations in the X-cube model, as the excitations created by each ${\cal M}_{\sigma_i}$ cancel out. The prism membrane operator counts the total number of $\fe_x$ and $\fe_y$ excitations inside the prism, modulo two, as can be seen from
\begin{equation}
{\cal M}_{{\rm prism}} = \prod_{i \in {\rm prism}} {\cal A}^z_i \text{.}
\end{equation}
Since ${\cal A}^i_z = -1$ acting on a state with an $\fe^i_x$ or $\fe^i_y$ excitation, the eigenvalue of ${\cal M}_{{\rm prism}}$ is 1 when an even number of such excitations are inside the prism, and $-1$ for an odd number.

The operator ${\cal M}_{{\rm prism}}$ can be viewed as effecting a process where two $\fm$ fractons are brought around the perimeter of the top and bottom faces of the prism, as shown in Fig.~\ref{fig:braiding_x_cube}a. Thinking in terms of the underlying toric code degrees of freedom, this corresponds to braiding a stack of $xy$-plane $m_P$ particles around the prism (Fig.~\ref{fig:braiding_x_cube}b), which results in a statistical phase of $\pi$ with any $xy$-plane $e_P$ particles contained inside. Each $\fe_x$ and $\fe_y$ particle is a bound state of an $xy$-plane $e_P$ particle with another $e_P$ particle, and contributes a phase of $\pi$. On the other hand, $\fe_z$ excitations are bound states of $yz$-plane and $xz$-plane $e_P$ particles, and do not contribute to the statistical phase.  Therefore this picture recovers the properties of ${\cal M}_{{\rm prism}}$ in the X-cube model deduced above.

\subsection{Logical operators and ground state degeneracy}
\label{sec:gsd} 

We now proceed to derive the ground state degeneracy (GSD) of the X-cube model on an $L\times L \times L$ three-torus $T^3$ (the calculation for other spatial topologies proceeds in a similar way). This result has been obtained previously by more rigorous methods for odd $L$.\cite{vijay2016fracton} Our approach relates the ground state degeneracy of the X-cube model to the underlying toric code degrees of freedom. 

To determine the ground state degeneracy, we need to count the number of independent logical operators in the theory. We first review this for the $d=2$ toric code on the two-torus $T^2$, where logical operators correspond to distinct ways to thread excitations around non-contractible cycles. We consider threading $m$ particles; we could just as well thread $e$ particles instead. We let $\gamma_x$ and $\gamma_y$ be paths winding around the different cycles of $T^2$, then $S_m(\gamma_x)$ and $S_m(\gamma_y)$ are string operators threading $m$ particles around the torus. It is simple to show that they are linearly independent. Thus, these two operators form a complete set of independent, commuting logical operators, and their eigenvalues completely label the ground state space, which has degeneracy ${\rm GSD} = 2^2 = 4$.

Now we consider the X-cube model, with $T^3$ topology obtained by enforcing periodic boundary conditions along each direction of the stack of toric codes, so that each plane in the stack has the topology of $T^2$. We count the number of distinct ways to thread magnetic $\cM_\sigma$ membranes through the non-contractible cycles of $T^3$. (We could just as well construct logical operators by threading $\fe$ particles around non-contractible cycles.)

A generic membrane operator ${\cal M}_\sigma$ can be constructed out of a stack of ${\cal S}_m(\gamma)$ string operators, where each path $\gamma$ lies in a single $\{100\}$ plane, which has the topology of a 2-torus. Therefore, in order to count independent logical operators, it is enough to consider the operators ${\cal S}_m(\gamma)$, where $\gamma$ winds around a non-contractible loop in its associated plane. Therefore we are led to consider logical operators that correspond to threading $m$ particles, which are really bound states of two $\fm$ fractons, around the torus. Each plane in the stack has two such independent logical operators, one for each of the independent non-contractible loops on the 2-torus, giving a contribution of $2^2$ to the GSD. Since there are a total of $3L$ planes in the stack, we obtain a total of $6L$ logical operators. 

However, these $6L$ logical operators are not all independent. To see this, let $(\mu,n)$ denote the plane normal to the $\mu$ direction and with $\mu$-coordinate $n$, and let $\gamma_{(\mu,n)}^\nu$ denote the path lying within the plane $(\mu,n)$ and passing around the non-contractible loop in the $\nu \neq \mu$ direction. Then we observe that for all $\mu \neq \nu$, we have the relation 
\be \label{linear_relns} \prod_{0\leq n < L} {\cal S}_m(\gamma_{(\mu,n)}^\nu) = \prod_{0\leq n < L} {\cal S}_m(\gamma^\mu_{(\nu,n)}) = {\cal M}_{\mu \nu},\ee
where ${\cal M}_{\mu \nu}$ is a membrane operator where the surface $\sigma$ covers an entire $\mu \nu$-plane. We have thus found one relation among the $6L$ logical operators for each (unordered) pair $\mu, \nu$ with $\mu \neq \nu$. There are three such pairs, and these three pairs exhaust the linear relations among the $6L$ logical operators
meaning the number of independent logical operators is $6L - 3$, a result we verified numerically using the methods of Appendix~\ref{app:fcc}. This results in a GSD of 
\be \log_2({\rm GSD}_{XC}) = 6L - 3,\ee
in agreement with the result obtained by commutative algebra methods in Ref.~\onlinecite{vijay2016fracton} for odd $L$. The sub-extensive $6L$ contribution is the GSD of $3L$ decoupled toric codes, and it is somewhat remarkable that the only modification of the GSD caused by the $J_z$ coupling term is a constant subleading correction which is independent of the number of toric codes in the stack. The analysis here is less rigorous than that of Ref.~\onlinecite{vijay2016fracton} because, in principle, we could have missed additional logical operators independent from the ones already listed. However, it is straightforward to check these results numerically for reasonably small values of $L$ as described in Appendix~\ref{app:fcc}. Numerical counting of the X-cube model gives the same number of independent stabilizers as above. Therefore, above $6L-3$ logical operators indeed form a complete set. 

\subsection{3d toric code from interlayer pair charge condensation of toric code layers}
\label{MCc}

As discussed at the end of Sec.~\ref{SC}, layers of toric codes interacting via $XX$ coupling result in the $d=3$ toric code model in the strong coupling limit ($J_x \to \infty$). Just like the case of $ZZ$ coupling, where the X-cube model results at strong coupling, there is a corresponding intermediate coupling picture that allows us to understand the excitations of the $d=3$ toric code in terms of the excitations of the $d=2$ toric code layers.

Acting with the coupling term $X^{\mu_1}_{\ell} X^{\mu_2}_{\ell}$ on the link $\ell$ creates a pair $e_1 e_2$ of $e$-particles at each end of $\ell$. Here $e_1$ and $e_2$ are the two-dimensional $e$-particles residing in the two perpendicular layers containing $\ell$. The composite $e_1 e_2$ is a one-dimensional particle that is expected to condense above a critical value of $J_x$ (recall that a $d$-dimensional particle is one that is constrained to move along a $d$-dimensional manifold).

In the presence of such a condensate, when $e_1$ approaches the intersection of the two planes, it can be converted into $e_2$ and move into the other plane. Therefore any two-dimensional $e$ particle in the lattice can convert to any other two-dimensional $e$ particle. Only a single type of $e$ particle remains as an independent excitation, and it is a three-dimensional particle. This is the point charge excitation (violation of the vertex term) of the $d=3$ toric code.

It is easy to see that the $e_1 e_2$ condensate confines single two-dimensional $m$ excitations. However, closed p-string configurations of $m$ particles have trivial statistics with the condensate. This follows from the analysis of Sec.~\ref{MCa}, upon noting that the $e_1 e_2$ particles forming the condensate here are the same as the one-dimensional $\fe$ excitations of the X-cube model, which also arise as $e_1 e_2$ composites. Those excitations are shown in Sec.~\ref{MCa} to have trivial statistics with closed p-strings. Here, this means that while single $m$ excitations are confined, closed p-strings of $m$ particles remain as deconfined excitations, and form the flux line excitations of the $d=3$ toric code.

\section{p-Membrane condensation: FCC model from coupled X-cube models}
\label{CXC} 

Above, we obtained the fracton topological order of the X-cube model from toric code layers via p-string condensation.  This raises the possibility of related condensation mechanisms that also lead to interesting fracton topological orders.  One option is to condense two-dimensional membranes built from particle excitations, or p-membranes.  Here, we describe a realization of p-membrane condensation in a system of four coupled X-cube models. We obtain a new exactly solvable model dubbed the Four Color Cube (FCC) model, which also happens to have a face centered cubic (fcc) lattice structure.  The p-membranes forming the condensate are composed of the one-dimensional $\fe$ particle excitations of the underlying X-cube models.  After discussing the coupling of X-cube models to obtain the FCC model, we discuss its properties and show that it possesses fracton topological order distinct from that of the X-cube model.

\begin{figure}[h]
\includegraphics[width=.7\columnwidth]{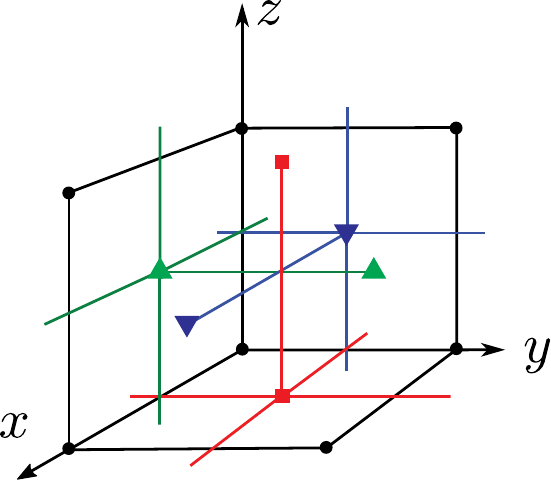}
\caption{(Color online) Lattice geometry of the coupled X-cube and FCC models, which are defined on four interpenetrating simple cubic lattices that we label by the colors black, red, green and blue. Dots, squares, up triangles and down triangles indicate the vertices of the four simple cubic lattices. Simple cubic lattice links intersect in mutually perpendicular triples of three different colors; for example, red, green and blue links intersect at centers of black cubes, as shown.  \label{fig:fcc}}
\end{figure}

The model is defined on four interpenetrating simple cubic lattices that we label by the colors black ($k$), red ($r$), green ($g$) and blue ($b$).  The geometry is shown in Fig.~\ref{fig:fcc}.  One way to understand the lattice geometry is to start with the black cubic lattice, and observe that there are three different orientations of plaquettes.  For each orientation, the plaquette centers form a simple cubic lattice, and these lattices are colored red ($xy$ plaquettes), green ($xz$ plaquettes) and blue ($yz$ plaquettes).  Taken together, all the vertices form a fcc lattice.  

We place an Ising spin on each link of a given color, for a total of 12 spins per simple cubic unit cell.  We use the term site (as opposed to vertex) to refer to the locations where spins reside.  Each site is an intersection of three mutually perpendicular links of three different colors, and we label sites by $\ell$. There are thus three spins located at $\ell$, with Pauli operators $X_\ell^w$, $Z_\ell^w$, where $w$ is any of the three colors located at $\ell$.  Sites with colors $w_1, w_2, w_3$ are located at cube centers of the $w_4$-colored lattice.  The sites also form a fcc lattice, so we can view our model as a fcc lattice spin system with three Ising spins per site.

The Hamiltonian of coupled X-cube models is
\begin{equation}
H_{CXC} = \sum_{w = k,r,g,b} H^w_{XC} - h \sum_{\ell} X^{w_1}_\ell X^{w_2}_\ell X^{w_3}_\ell \text{,}
\end{equation}
where the first term is simply four decoupled X-cube Hamiltonians on the four cubic lattices, and the second term couples the different colors, where $w_1, w_2$ and $w_3$ are the three different colors at $\ell$.  The Hamiltonian of the color $w$ X-cube model is written
\begin{equation}
H^w_{XC} = - \sum_{i \in w} \sum_{\mu = x,y,z} A^{\mu}_i - \sum_{c \in w} B_c \text{,}
\end{equation}
where we have taken the coefficients of the two terms to be equal, the first sum is over all vertices in the color $w$ cubic lattice, and the second sum is over all cubes in the color $w$ lattice.  Since the notation we are using here differs slightly from the previous sections, we again give the form of the stabilizers.  The $Z$-stabilizers are 
\begin{equation}
A^{\mu}_i = \prod_{ij \perp \mu} Z_{i j} \text{,}
\end{equation}
where $j$ is a vertex adjacent to $i$ in the lattice of the same color, and the product is over links $ij$ perpendicular to $\mu$.  The $X$-stabilizers are
\begin{equation}
B_c = \prod_{\ell \in c} X^w_{\ell} \text{,}
\end{equation}
where the product is over edges of the cube $c$, and the color $w$ is specified by the choice of $c$.

The Hamiltonian $H_{CXC}$ has the full translational symmetry of the fcc lattice, provided translations are accompanied by certain spin rotations that correspond to permuting the colors, \emph{i.e.} translation acts in a ``spin-orbit coupled'' manner.   Under translation by a Bravais lattice basis vector $\ba_1, \ba_2, \ba_3$, we make the following permutations of colors
\begin{eqnarray}
\ba_1 &:&   k \leftrightarrow b, r \leftrightarrow g \\
\ba_2 &:&  k \leftrightarrow g, r \leftrightarrow b \\
\ba_3 &:& k \leftrightarrow r, g \leftrightarrow b \text{,}
\end{eqnarray}
where the basis vectors are
\begin{eqnarray}
\ba_1 &=& \Big( 0, \frac{1}{2}, \frac{1}{2} \Big) \\
\ba_2 &=& \Big( \frac{1}{2}, 0, \frac{1}{2} \Big) \\
\ba_3 &=& \Big( \frac{1}{2}, \frac{1}{2}, 0 \Big) \text{.}
\end{eqnarray}
Here and elsewhere we set the distance between neighboring vertices of the same color to one.

\begin{figure}[h]
\includegraphics[width=.3\textwidth]{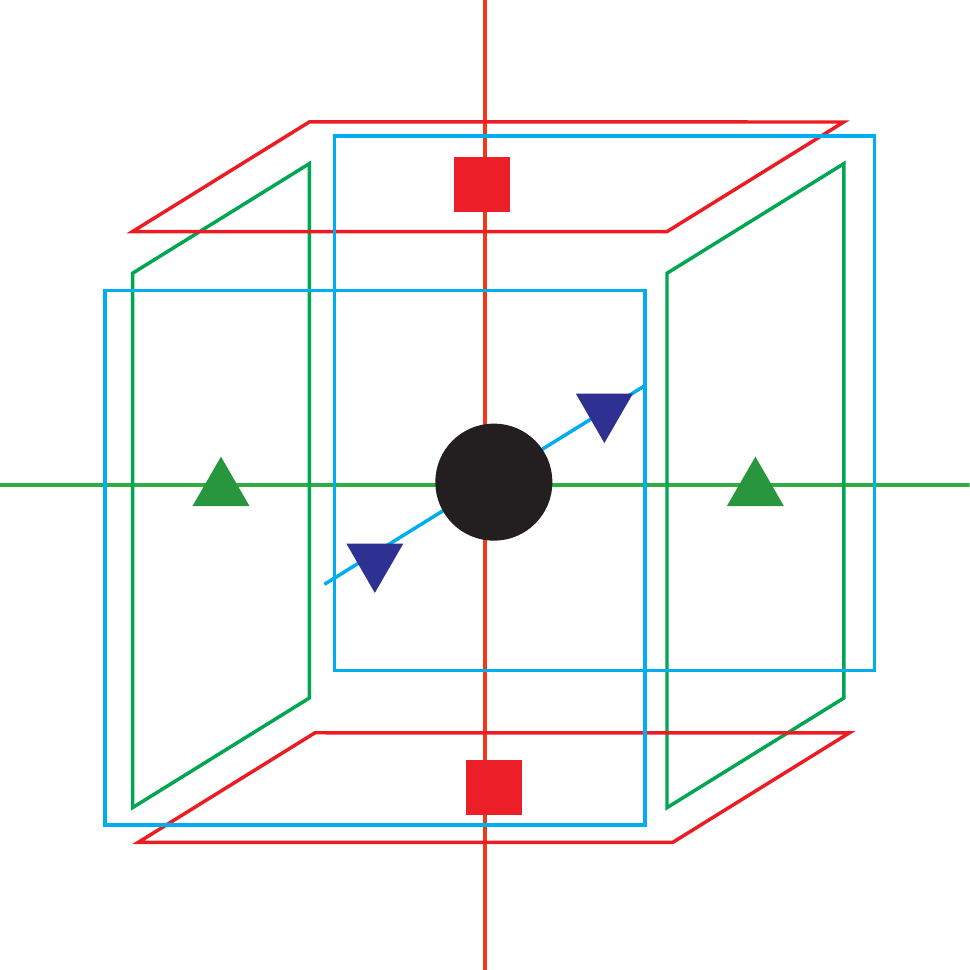}
\caption{(Color online) The p-membrane condensation in FCC model. The big black dot denotes one of the cubes in the black X-cube model. It is surrounded by plaquettes of the other three colors representing the planes perpendicular to the motion of $d=1$ particles, generated at vertices of different cubic lattices denoted by different shapes as shown in Fig. (\ref{fig:fcc}).  \label{fig:p_membrane}}
\end{figure}

Before proceeding to analyze $H_{CXC}$, we first motivate the form of the coupling.  The coupling term creates two-dimensional p-membrane objects that are expected to condense for $h$ sufficiently large.  To see this, we consider the effect of acting with $H^{{\rm int}}_{\ell} = X^{w_1}_\ell X^{w_2}_\ell X^{w_3}_\ell$ in the decoupled ($h = 0$) limit.  Each $X^w_\ell$ operator creates a pair of one-dimensional $\fe$ particles at the two endpoints of the link $\ell$, in the color $w$ X-cube model.  Therefore, $H^{{\rm int}}_{\ell}$ creates six such excitations located at the vertices of an octahedron centered at the site $\ell$ (Fig.~\ref{fig:p_membrane}).  It is natural to view each $\fe$ particle as a square plaquette, as illustrated in Fig.~\ref{fig:p_membrane}, where the plane of the square represents the directions in which $\fe$ cannot move. The six squares join together to form a closed cube surrounding $\ell$, so that we can view $H^{{\rm int}}_\ell$ as creating a small closed p-membrane.  It is then natural to conjecture that the physics of the large-$h$ limit can be understood in terms of condensation of p-membranes.  This is directly analogous to p-string condensation, where we represented two-dimensional $m$ particles as line segments indicating the direction in which the particle cannot move, and these line segments join together into closed p-loops.  In both cases, the dimension of the object condensing (p-string or p-membrane) is the co-dimension of the space in which the constituent particle excitations move.

We proceed with our analysis of $H_{CXC}$ by considering the strong coupling ($h \to \infty$) limit, where we obtain the exactly solvable FCC model in degenerate perturbation theory.  For a single site, in the $h \to \infty$ limit we have the constraint
\begin{equation}
\label{eqn:fcc-constraint}
X^{w_1}_\ell X^{w_2}_\ell X^{w_3}_\ell = 1 \text{,}
\end{equation}
which defines a low-energy Hilbert space of two effective Ising spins.  Rather than solving the constraint, \emph{e.g.} by eliminating one of the $X^w_{\ell}$ operators, we find it convenient to work in the constrained Hilbert space of three spins.  Because the constraint is on-site, this is purely a matter of convenience.  An arbitrary single-site operator is built from sums and products of $X^w_{\ell}$ and $Z^{w_1}_{\ell} Z^{w_2}_{\ell}$; single $Z^w_{\ell}$ operators do not commute with the constraint in Eq. (\ref{eqn:fcc-constraint}). 

To understand the large-$h$ limit, we need to carry out degenerate perturbation theory.  As discussed in Appendix~\ref{app:perturbation-theory}, the necessary calculation is essentially the same as that in the large-$J_z$ limit of coupled toric codes described in Sec.~\ref{SC} and in Appendix~\ref{app:perturbation-theory}.  At sixth order in perturbation theory we obtain the FCC model,
\begin{equation}
H_{FCC} = - \sum_{c} B_c - K \sum_{c} A_c \text{,}
\end{equation}
where $K$ is a positive constant proportional to  $1/h^5$.  The sums are over all cubes $c$, in all four cubic lattices.  All the terms in $H_{FCC}$ are stabilizers, or products of stabilizers, of the underlying X-cube models, so that any two terms commute and the model is exactly solvable.

The first term in $H_{FCC}$ is simply the projection of the $X$-stabilizer terms of the X-cube models into the low-energy subspace.  Since these terms commute with the constraint, their form is unaffected, although we do need to keep in mind that the underlying $X^w_{\ell}$ operators now obey the constraint Eq.~(\ref{eqn:fcc-constraint}).

\begin{figure}[h]
\includegraphics[width=.7\columnwidth]{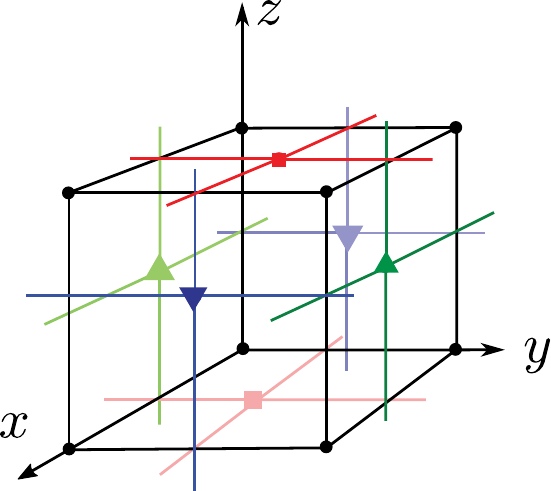}
\caption{(Color online) Illustration of a FCC model $A_c$ operator on a black cube. The operator is a product of $A^{\mu}_i$ over the cube's six faces shown by the red, green and blue links, whose vertices are represented by corresponding shapes shown in Fig. (\ref{fig:fcc}). For each edge of the cube, two $Z^w_{\ell}$  operators contribute to $A_c$.\label{fig:ac}}
\end{figure}

The operators $A_c$ in the second term, which are the $Z$-stabilizers of the FCC model, are obtained as the smallest non-constant products of X-cube $Z$-stabilizers ($A^{\mu}_i$'s) that commute with the constraint (Eq. (\ref{eqn:fcc-constraint})).  To understand the form of $A_c$, we observe that each vertex term $A^{\mu}_i$ in one of the underlying X-cube models lies on a face $f$ of a cube $c$, where $c$ and the vertex $i$ have different colors, as shown in Fig.~\ref{fig:ac}. We thus write $A_f \equiv A^{\mu}_i$.  Then we obtain $A_c$ by taking a product over the six faces of $c$,
\begin{equation}
A_c = \prod_{f \in c} A_f \text{.}
\end{equation}
As illustrated in Fig.~\ref{fig:ac}, $A_c$ has two $Z^{w}_{\ell}$ operators on each edge of $c$, so it commutes with the constraint.

The FCC model obeys an electric-magnetic self-duality that we will exploit in our analysis.  To expose the duality, we define new Pauli operators
${\cal Z}^w_{\ell}$ and ${\cal X}^w_{\ell}$ by
\begin{eqnarray}
{\cal X}^{w_1}_{\ell} &=& Z^{w_2}_{\ell} Z^{w_3}_{\ell} \\
{\cal Z}^{w_1}_{\ell} {\cal Z}^{w_2}_{\ell} &=& X^{w_3}_{\ell} \text{,}
\end{eqnarray}
where $w_1, w_2, w_3$ are the three distinct colors at $\ell$.  These new operators obey the same algebraic relations as the constraint
${\cal X}^{w_1}_{\ell} {\cal X}^{w_2}_{\ell} {\cal X}^{w_3}_{\ell}  = 1$.  If we set $K = 1$ in $H_{FCC}$, this change of variables becomes a symmetry of the model, where the $A_c$ and $B_c$ stabilizers are exchanged.

The electric-magnetic self-duality appears quite surprising, given that the starting point of coupled X-cube models has no such property.  However, it can be rationalized by going back to the construction of the underlying X-cube models from toric code layers.  In the limit of decoupled toric codes, there is of course an electric-magnetic self-duality.  To obtain X-cube models, we condense p-strings of $m$ particles, and then obtain the FCC model by condensing p-membranes of $\fe$ particles.  Since condensations occur in both the electric and magnetic sectors, it is reasonable that electric-magnetic self-duality can be restored in the FCC model.  This suggests that there might be a manifestly self-dual route directly from toric code layers to the FCC model, bypassing the intermediate step of X-cube models; we leave exploration of this possibility to future work.

Now we turn to an analysis of the FCC model. The first property to establish is that the model has topological order. That is, we would like to argue that there is a non-trivial ground state degeneracy on the 3-torus, and that the degenerate ground states cannot be distinguished by local measurements. We argue that this is the case in Appendix~\ref{app:fcc}, where we find that the ground state degeneracy ${\rm GSD}$ on a $L \times L \times L$ torus satisifies
\begin{equation}
\operatorname{log}_2 {\rm GSD} = 32 L - 24 \text{.}
\end{equation}
We note that this is \emph{not} simply four times the result for a single X-cube model, which is $4 \times (6 L - 3) = 24 L - 12$.

\begin{figure}[h]
\includegraphics[width=.9\columnwidth]{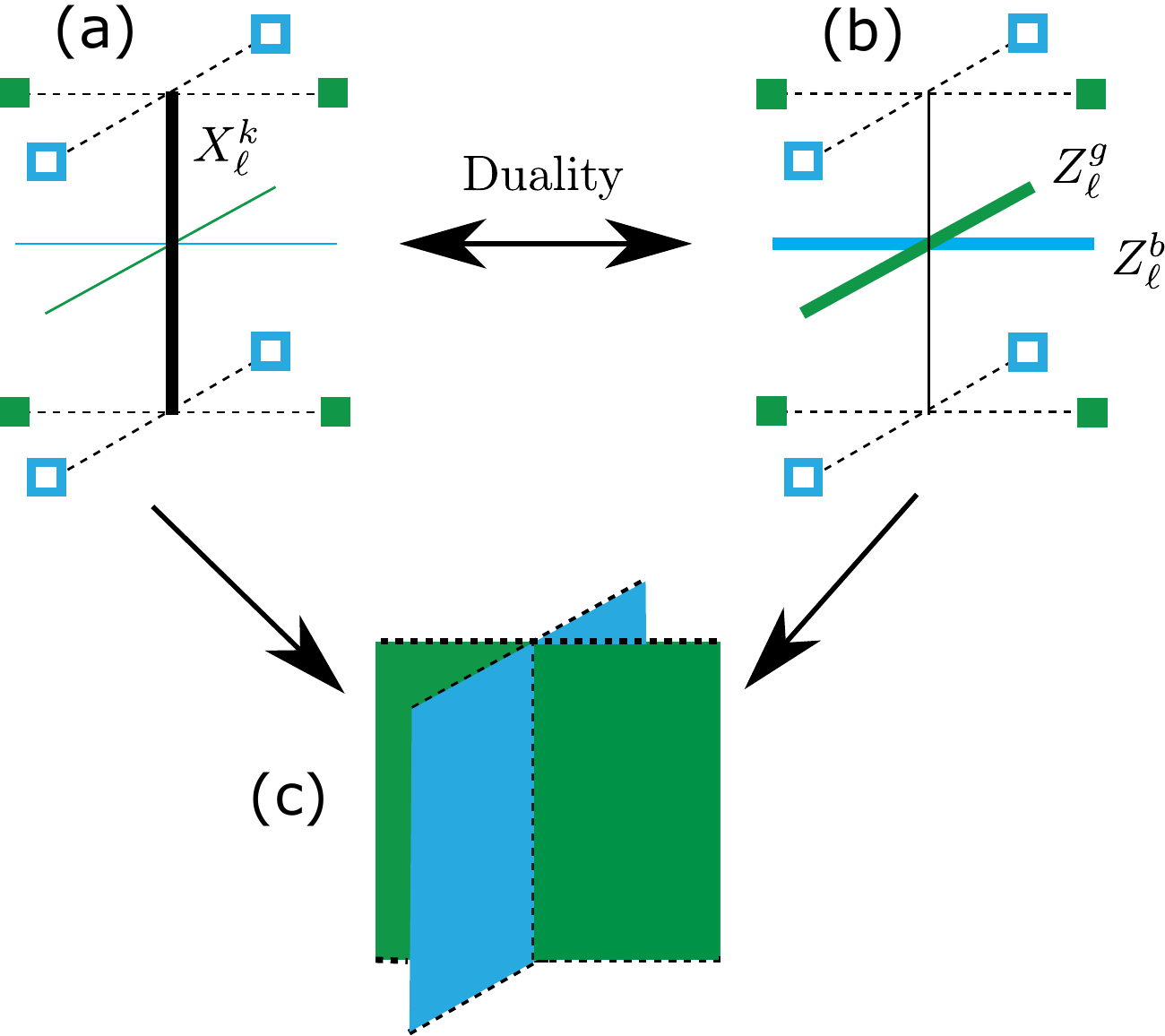}
\caption{(Color online) (a) Acting with $X^k_{\ell}$ on the thick black link (pointing in the $z$ direction) creates eight $A_c = -1$ excitations, four of which are blue (open squares) and four green (solid squares). (b) $X^k_{\ell}$ is dual to $Z^g_{\ell} Z^b_{\ell}$, where each $Z$ operator creates four $B_c = -1$ excitations of the same color, again for a total of eight excitations, which illustrates the self-duality. The four excitations created by each $Z^w_{\ell}$ are four fractons in one of the underlying X-cube models.
(c) In both cases, the eight excitations can be viewed as created at corners of two perpendicular membrane operators.\label{fig:xoperator}}
\end{figure}

Next, we discuss the excitations of the FCC model. As illustrated in Fig.~\ref{fig:xoperator}a, acting with $X^{w_1}_{\ell}$ on a ground state creates eight $A_c = -1$ excitations. $X^{w_1}_{\ell}$ is dual to $Z^{w_2}_{\ell} Z^{w_3}_{\ell}$, which also creates eight $B_c = -1$ excitations, illustrating the self-duality (Fig.~\ref{fig:xoperator}b). Each $Z^w_{\ell}$ operator can be thought of as creating four fractons in the underlying color-$w$ X-cube model, which are created at corners of a membrane perpendicular to the color-$w$ link $\ell$. Therefore the eight excitations created by $Z^{w_2}_{\ell} Z^{w_3}_{\ell}$ can be viewed as created at corners of two perpendicular membrane operators, as shown in Fig.~\ref{fig:xoperator}c. By self-duality, the same picture holds for the eight $A_c = -1$ excitations created by $X^{w_1}_{\ell}$.

\begin{figure}[h]
\includegraphics[width=.9\columnwidth]{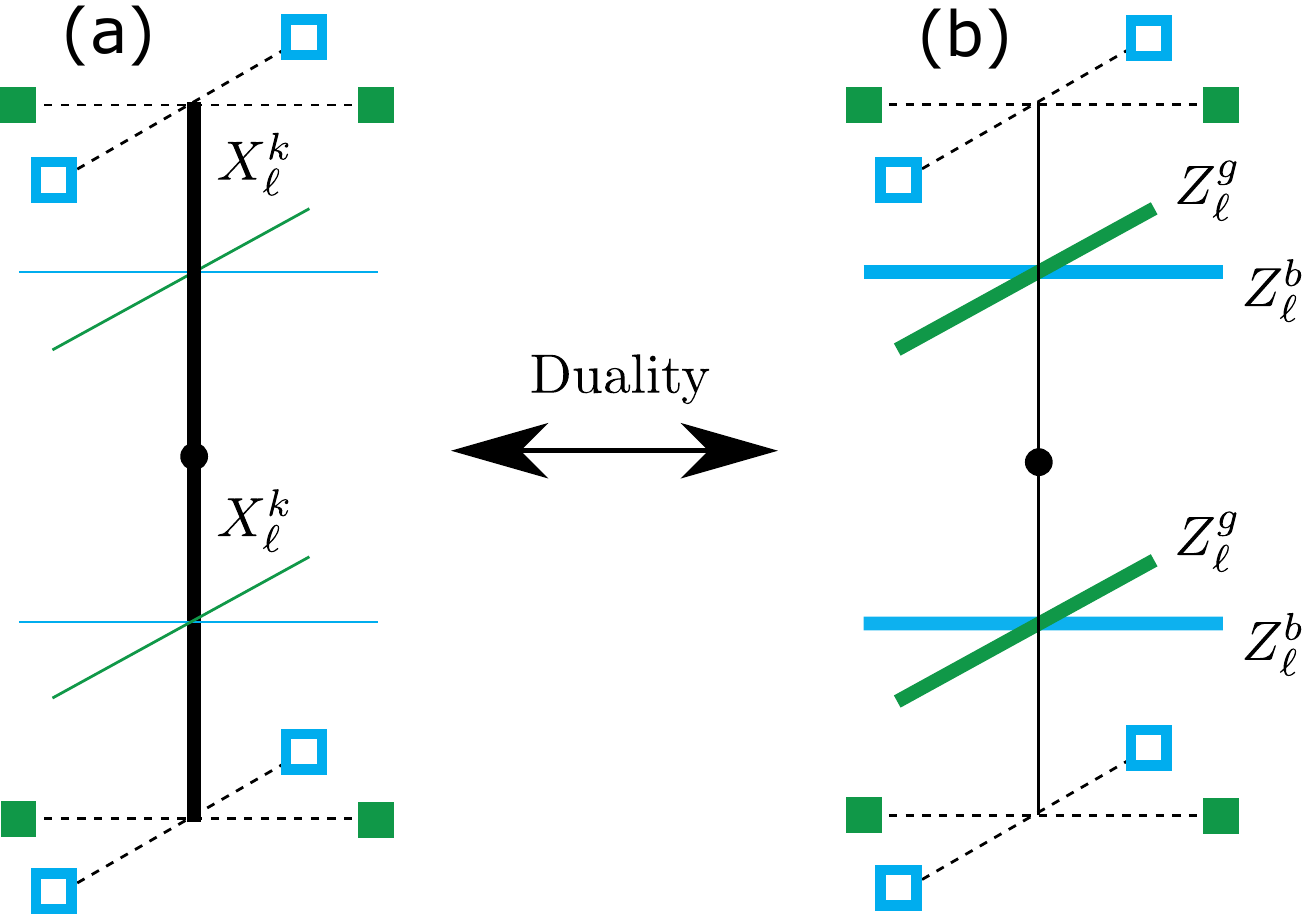}
\caption{(Color online) (a) Acting with a product of $X^k_{\ell}$ along a straight line oriented along the $z$-axis creates one-dimensional particle excitations at the end of the line, each composed of four $A_c = -1$ excitations (open/solid squares). These excitations are the remnants of one-dimensional $\fe$ excitations in the black X-cube model.  (b) Electric-magnetic duality shows that acting with a product of $Z^g_{\ell} Z^b_{\ell}$ along a line in the $z$-direction also creates one-dimensional particle excitations. These one-dimensional excitations can be thought of as bound states of two-fracton two-dimensional particles from green and blue X-cube models. The blue bound state moves in a $yz$ plane, while the green bound state moves in a $xz$ plane, so the bound state of both of them is constrained to move in the $z$-direction. \label{fig:1d_particles}}
\end{figure}

Acting with a product of $X^w_{\ell}$ along a straight line creates one-dimensional particle excitations at ends of the line, as illustrated in Fig.~\ref{fig:1d_particles}a. These particles are made up of four $A_c = -1$ excitations, and are remnants of the one-dimensional $\fe$ excitations of the underlying X-cube models. Electric-magnetic self-duality shows there are corresponding one-dimensional particles made up of four $B_c = -1$ excitations, as shown in Fig.~\ref{fig:1d_particles}b.

\begin{figure}[h]
\includegraphics[width=.9\columnwidth]{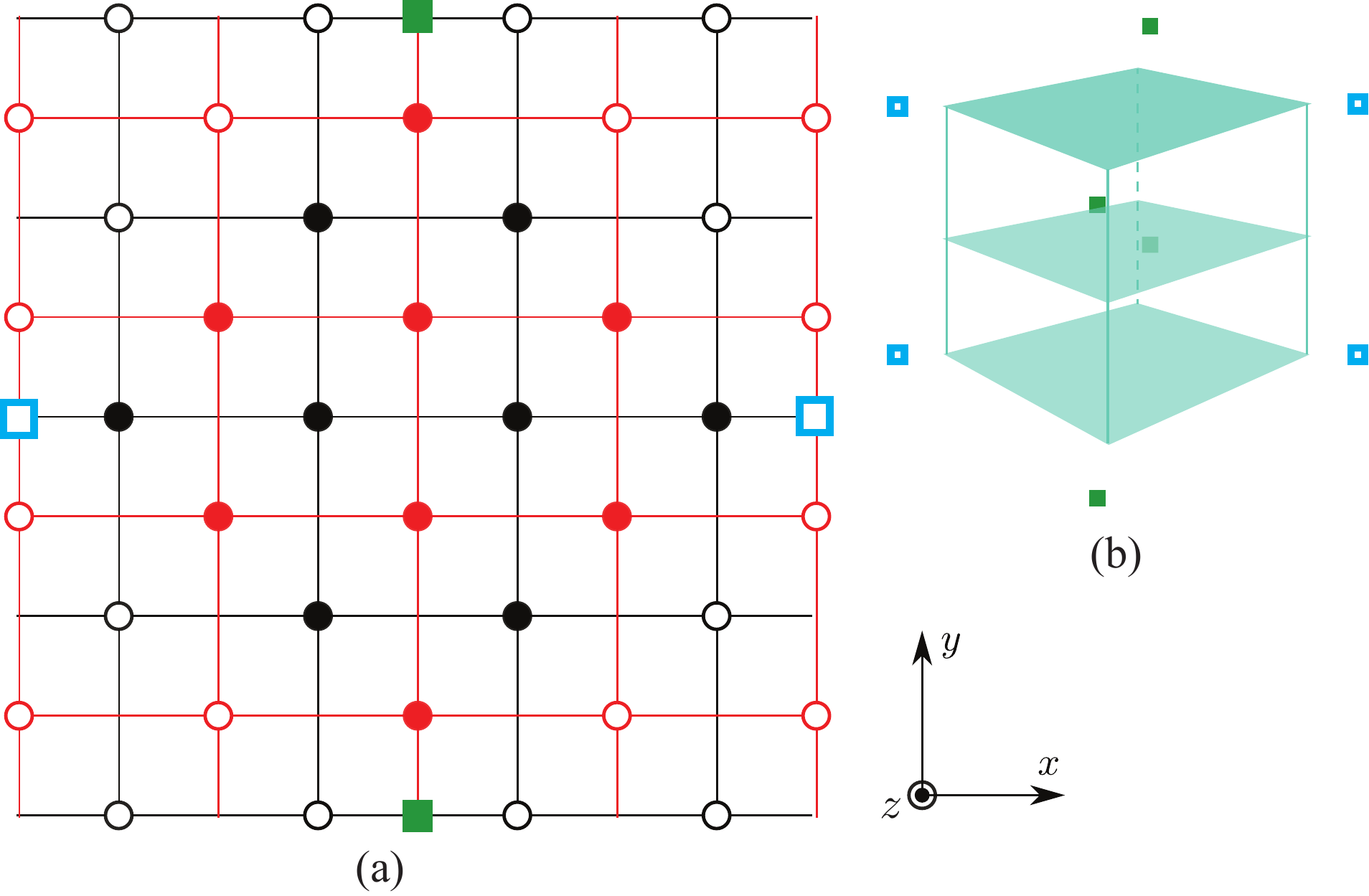}
\caption{(Color online) (a) Diamond pattern in a $(001)$ plane that comprises a single layer of the skyscraper operator. Each circle represents a black or red link pointing in the $z$-direction, and a product of $X^w_{\ell}$ is taken over the filled circles. $A_c = -1$ excitations are created at the corners at the top and bottom of the skyscraper, as indicated by the blue open squares and green solid squares. The filled circles also indicate the positions of $\fe$ particles created at the top and bottom of the skyscraper, when we act with the skyscraper operator in the limit of decoupled X-cube models. (b) A three-dimensional view of the skyscraper operator. Each green layer represents a diamond pattern as shown in (a), and the locations of the eight $A_c = -1$ excitations are shown by blue squares.  \label{fig:fcc_fractons}}
\end{figure}

Isolated $A_c = -1$ excitations can be created by a ``skyscraper operator,'' which is a product of $X^w_{\ell}$ over a pattern resembling a skyscraper, which we illustrate with a particular example. The skyscraper is formed from a stack of $(001)$ planes, where on each plane we act with a product of $X^w_{\ell}$ over a diamond pattern of black and red links pointing in the $z$-direction, as shown in Fig.~\ref{fig:fcc_fractons}. The skyscraper operator thus consists of black and red string operators for the one-dimensional particles shown in Fig.~\ref{fig:1d_particles}a, with the strings forming a diamond pattern when the skyscraper is viewed from ``above'' in the $z$-direction. At the top and bottom of the skyscraper, $A_c = -1$ excitations are created at the corners as shown in Fig.~\ref{fig:fcc_fractons}. Electric-magnetic self-duality implies that isolated $B_c = -1$ excitations can be created in a corresponding manner.

If we act with the skyscraper operator in the limit of decoupled X-cube models, we create p-membranes of black and red $\fe$ particles at the top and bottom of the skyscraper, in the pattern indicated by filled circles in Fig.~\ref{fig:fcc_fractons}a.  In the FCC model, the only excitations lie at the corners, which illustrates that the two-dimensional ``bulk'' of the p-membrane, and its one-dimensional edges, are condensed.

Naively, the skyscraper construction suggests that isolated $A_c = -1$ excitations are created at corners of a \emph{volume} operator, with support over the interior of a solid three-dimensional region. However, we can act with $X$-stabilizers to ``hollow out'' the inside of the skyscraper. Specifically, we act with products of black and red $B_c$ operators over non-overlapping vertical columns, which cancel out the string operators on the inside of the skyscraper, and introduce new $X^w_{\ell}$ operators on the top and bottom faces. This shows that $A_c = -1$ excitations are created at corners of a membrane operator supported over a two-dimensional region, the boundary of the skyscraper. However, the resulting object appears complicated geometrically, and so far we have found the skyscraper construction more useful for visualization purposes.

We now argue that isolated $A_c = -1$ and $B_c = -1$ excitations are immobile fractons, using statistical properties of these excitations. Self-duality allows us to focus on $B_c = -1$ excitations for convenience. Suppose that a single isolated $B_c = -1$ excitation is contained in some volume. We can detect this excitation by acting with the operator
\begin{equation}
{\cal B} = \prod_{c' \in \text{prism} } B_{c'} \text{,}
\end{equation}
where ``prism'' is a rectangular prism containing $c$, and includes only cubes of the same color as $c$.  ${\cal B}$ has eigenvalue $1$ acting on the ground state, and eigenvalue $-1$ on the state with the excitation. Most of the $X^w_{\ell}$ operators in the product defining ${\cal B}$ cancel out, and, indeed, ${\cal B}$ is a product of $X^w_{\ell}$ over the edges of the rectangular prism. That is, ${\cal B}$ is comprised of string operators for the one-dimensional particles of Fig.~\ref{fig:1d_particles}a on the edges of the prism.

\begin{figure}[h]
\includegraphics[width=.6\columnwidth]{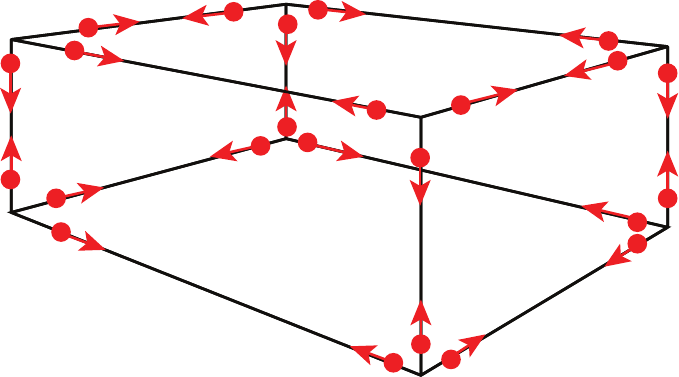}
\caption{(Color online) Acting with the operator ${\cal B}$ can be viewed as implementing a process where three one-dimensional particles are created at each corner of a rectangular prism, and these particles are then moved along edges to annihilate. This process has a non-trivial statistical phase of $\pi$ when a single $B_c = -1$ excitation is present inside the prism, where $c$ and the edges of the prism all belong to the same color cubic lattice. \label{fig:frame}}
\end{figure}

Acting with ${\cal B}$ effects a process where three one-dimensional particles are created at each corner of the prism, and are then brought together along edges to annihilate, as shown in Fig.~\ref{fig:frame}. This process can be used to remotely detect the $B_c = -1$ excitation contained inside the prism. This is only consistent if this excitation is a fracton, meaning that it cannot be moved by acting with any string operator. Indeed, if it could be transported in this way, acting with the string operator could move it out of the prism through one of the faces.
Since the size and shape of the ${\cal B}$ prism operator is arbitrary (as long as it contains the $B_c=-1$ excitation), ${\cal B}$ can always be chosen so that any putative string operator moving the excitation does not intersect the edges of ${\cal B}$, guaranteeing that the excitation can be trivially moved out of the prism. This is inconsistent with the fact that ${\cal B}$ can be used to remotely detect the excitation, and so the excitation must be a fracton.

\section{Semionic X-cube model}
\label{SXC} 

Interpreting the X-cube model in terms of a coupled layer construction allows us to generalize the model. In this section, we show how to generalize the construction to coupled layers of the double semion model so that the resulting ``semionic'' X-cube model has ``semionic'' 1D particles. In this section, in contrast with the discussion of the coupled-layer construction of the X-cube model, we will not be as concerned with the fine details of degenerate perturbation theory in the strong coupling limit.  In particular, we will not try to establish the absence of perturbations to the semionic X-cube model in sixth order perturbation theory.  Answering this question will not be important for our main purpose, which is to discover the semionic X-cube model, to understand some of its properties, and to understand how its excitations are related to those of the underlying double semion layers.

\begin{figure}[htbp]
\centering
\includegraphics[width=0.5\textwidth]{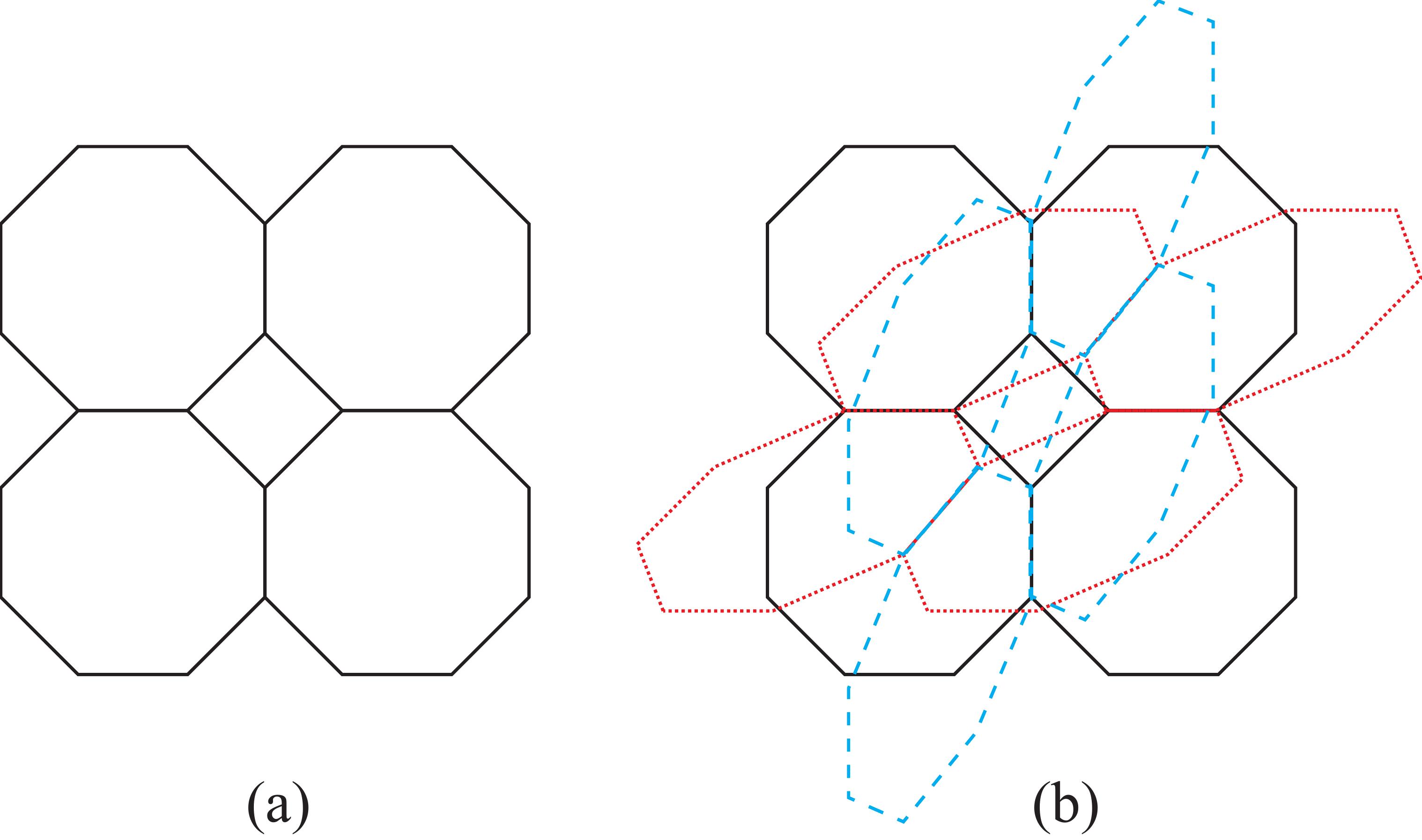}
\caption{(Color online) (a) A trivalent 2D lattice from decorating 2D square lattice with diamond shapes at each vertex. (b) Stacks of such trivalent lattices in $x$ (solid lines), $y$ (dash lines), and $z$ (dotted lines) planes; the edges in $x$, $y$ and $z$ directions overlap in pairs.}
\label{triv_stack}
\end{figure}

To construct the ``semionic'' version of the model, we will need to start from trivalent 2D lattices instead of the square lattices as discussed in previous sections. First, we demonstrate how to obtain from this starting point a model with the same topological order as the X-cube model. Consider the decorated square lattice as shown in Fig.\ref{triv_stack} (a). A small diamond shape is added at each vertex of the square lattice so that in the new lattice each vertex has degree three. 

To define a toric code on the decorated lattice, we put spins on all the edges and impose Hamiltonian terms as
\begin{equation}
\tilde{H}^{TC} = - \sum_{ \includegraphics[height=0.1in]{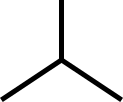}} \prod_{\ell \in \includegraphics[height=0.1in]{triv.png}} Z_\ell - \sum_{\includegraphics[height=0.1in]{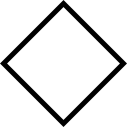}} \prod_{\ell \in \includegraphics[height=0.1in]{diam.png}} X_\ell - \sum_{\includegraphics[height=0.1in]{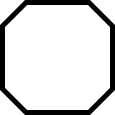}} \prod_{\ell \in \includegraphics[height=0.1in]{oct.png}} X_\ell 
\text{.}
\end{equation}
The topological order of this model remains the same as the square lattice toric code model \eqref{TC_sq}, with the only difference coming from the fact that there can be extra plaquette excitations in the diamonds which correspond to the same type of anyon as the octagon (originally square) plaquette excitations.  

Now we take three stacks of such 2D models in the $x$, $y$ and $z$ planes and couple them as shown in Fig.\ref{triv_stack} (b). The planes are positioned in such a way that edges in the $x$, $y$ and $z$ directions overlap in pairs and share two spins. The diagonal edges do not overlap. For overlapping edges, we denote Pauli operators by $X^{\mu}_{\ell}$ and $Z^{\mu}_{\ell}$, where $\mu$ gives the normal direction of the toric code layer containing the spin. We couple the layers by adding a $Z\otimes Z$ term on each pair of overlapping edges, as in our coupled-layer construction of the original X-cube model. The total Hamiltonian becomes
\begin{equation}
\tilde{H} = \sum_P \tilde{H}^{TC}_P - J_z \sum_{\ell \parallel x,y,z} Z^{\mu_1}_{\ell}  Z^{\mu_2}_{\ell} \text{.}
\end{equation}
where the latter sum is over edges aligned in the $x$, $y$ and $z$ directions.

Among the original Hamiltonian terms, the vertex terms and the plaquette terms on the diamonds commute with the coupling term while the plaquette terms on the octagons anticommute with the coupling. Therefore, if we pass to the strong coupling $J_z \rightarrow \infty$ limit, the octagon terms need to reorganize and combine into cubes. In the strong coupling limit, the two overlapping edges combine into one and there is effectively one Ising spin per edge. As in Sec.~\ref{SC}, we introduce ${\cal Z}_{\ell}$ and ${\cal X}_{\ell}$ Pauli operators for the effective Ising spins in the $J_z \to \infty$ limit, and the effective Hamiltonian becomes 
\begin{equation}
\tilde{H}_{XC} = - \sum_{ \includegraphics[height=0.1in]{triv.png}} \prod_{\ell \in \includegraphics[height=0.1in]{triv.png}} {\cal Z}_\ell - \sum_{\includegraphics[height=0.1in]{diam.png}} \prod_{\ell \in \includegraphics[height=0.1in]{diam.png}} {\cal X}_\ell - \sum_{\includegraphics[height=0.15in]{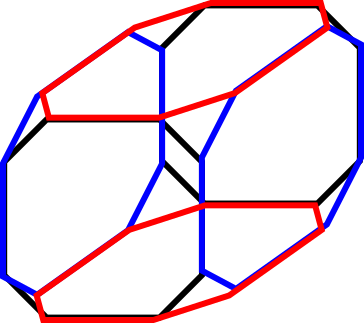}} \prod_{\ell \in \includegraphics[height=0.15in]{oct_cube.png}} {\cal X}_\ell 
\text{.}
\label{triv_XC}
\end{equation}

\begin{figure}[htbp]
\centering
\includegraphics[width=0.5\textwidth]{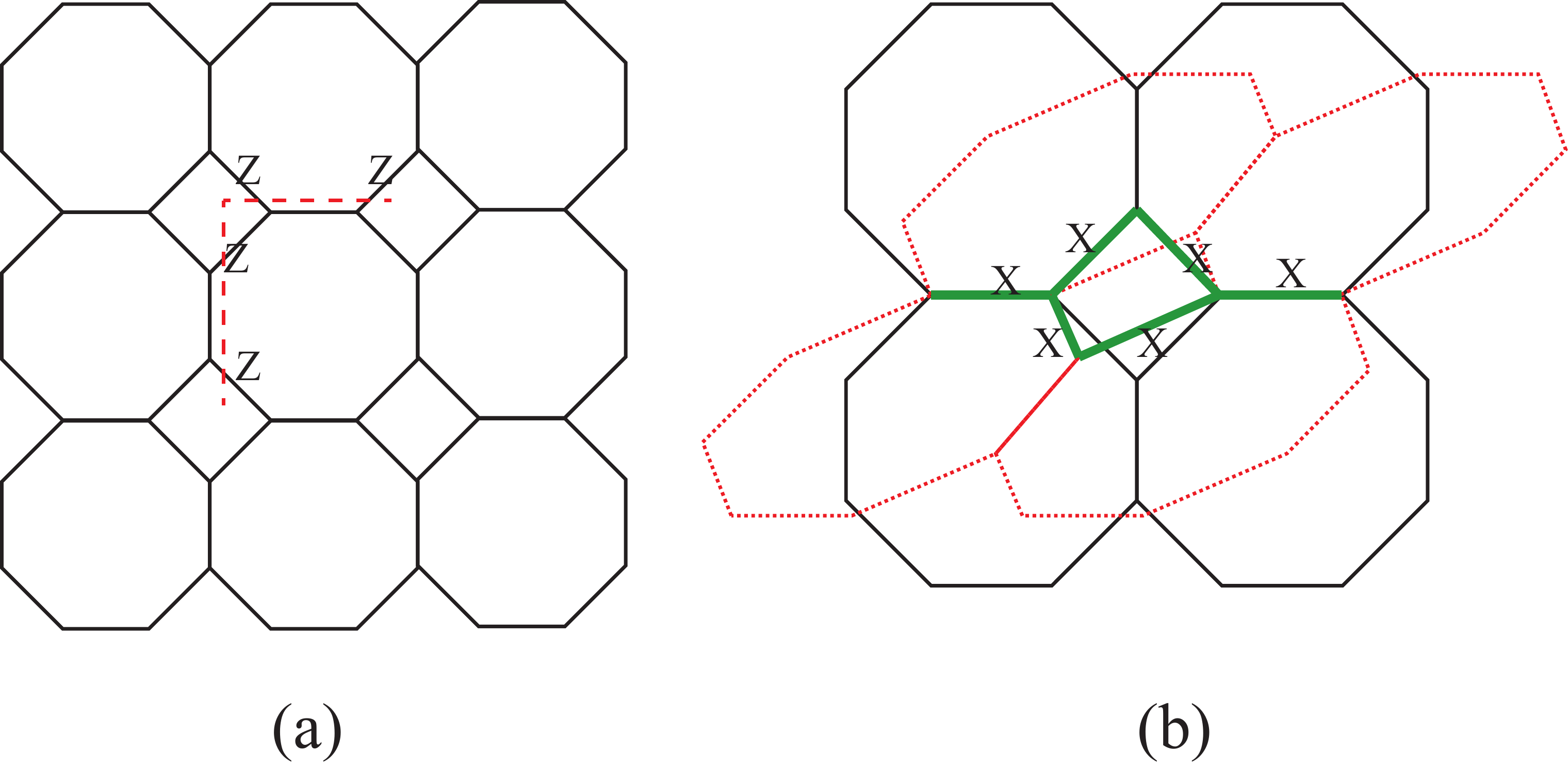}
\caption{(Color online) Illustration of the properties of excitations in the modified X-cube model on a decorated lattice. (a) Excitations in diamond plaquettes can move freely in 2D with a string of ${\cal Z}$ operators as shown; (b) Excitations at vertices can move in 1D with a string of ${\cal X}$ operators as shown.}
\label{triv_ex}
\end{figure}

Now we are in a position to examine the excitations of this model. If we apply ${\cal Z}_{\ell}$ for $\ell$ an edge in the $x$, $y$ and $z$ direction, we create four cube excitations, which can be separated to four corners of a membrane by subsequent action of ${\cal Z}_{\ell}$ operators on edges perpendicular to the membrane. These are the fracton excitations of the modified X-cube model. If we apply ${\cal Z}_{\ell}$ to one of the edges of the diamond plaquettes, we create two cube excitations together with a diamond plaquette excitation. This is saying that two fractons are equivalent to a diamond plaquette excitation. As the combination of two fractons can move freely in a two dimensional plane, so can the diamond plaquette excitation. This can be seen from Fig.\ref{triv_ex} (a) where a string of ${\cal Z}_{\ell}$ operators can move a diamond plaquette excitation around. 

If we apply ${\cal X}_{\ell}$ to one of the edges of the diamond plaquettes, we create one vertex excitation at each end of $\ell$. If we want to move these vertex excitations, we need to apply ${\cal X}_{\ell}$ to the edges in $x$, $y$ and $z$ directions, which creates \emph{two} vertex excitations at each end of $\ell$. Similarly to the original X-cube model, vertex excitations on intersecting planes move together, which restricts their motion to the intersection line. The only small difference is that when a pair of vertex excitations passes through a diamond plaquette, their path separates onto two planes before merging again, as shown in Fig.\ref{triv_ex} (b).

In this way, by starting from toric code layers on trivalent 2D lattices, we obtain a generalized X-cube model with the same topological content. 

To obtain the semion version of the X-cube model, we start with the double semion model\cite{levin05} on the decorated square lattice with Hamiltonian 
\begin{equation}
\begin{array}{ll}
\tilde{H}^{DS} = &  - \sum_{ \includegraphics[height=0.1in]{triv.png}} \prod_{\ell \in \includegraphics[height=0.1in]{triv.png}} Z_\ell - \sum_{\includegraphics[height=0.1in]{diam.png}} P_v\left(\prod_{\ell \in \includegraphics[height=0.1in]{diam.png}} X_\ell \prod_{\ell \in \includegraphics[height=0.2in]{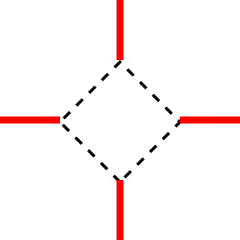}} S_\ell \right)\\
&-\sum_{\includegraphics[height=0.1in]{oct.png}} P_v\left(\prod_{\ell \in \includegraphics[height=0.1in]{oct.png}} X_\ell \prod_{\ell \in \includegraphics[height=0.2in]{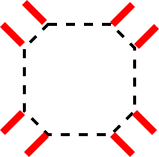}} S_\ell\right)
\text{.}
\end{array}
\end{equation}
where the second product in each plaquette term is over $S=\begin{pmatrix} 1 & 0\\ 0 & i\end{pmatrix}$ on all the legs pointing outward from the plaquette, and $P_v$ is a projector onto configurations satisfying the vertex term for all the vertices contained in the plaquette. This model is known to have the double semion topological order.\footnote{The lattice geometry differs from that considered in Ref.~\onlinecite{levin05}.  However, because the lattice is still trivalent, the analysis of Ref.~\onlinecite{levin05} goes through, and the model supports double semion topological order.} In particular, there are semion string operators which anti-commute with each other when they intersect.

Now we couple the layers, using the same coupling term as in the toric code case:
\begin{equation}
\tilde{H}^{s} = \sum_P \tilde{H}^{DS}_P - J_z \sum_{\ell} Z^{\mu_1}_{\ell}  Z^{\mu_2}_{\ell} \text{.}
\end{equation}
In the strong coupling limit, overlapping edges combine into one. The vertex and diamond plaquette terms commute with the $J_z$ term, and lead to terms of the same form in first-order perturbation theory. The octagon plaquette terms do not commute with the $J_z$ term, and a product of six such terms over the faces of a cube appears at 6th order in perturbation theory.  It is possible that other terms appear in the perturbation series, but we ignore them (see remarks at the beginning of this section). The resulting Hamiltonian takes a form very similar to Eq. \eqref{triv_XC}, except that the diamond plaquette terms and the octagon cube terms are supplemented with a product of $S$ operators over outward pointing edges, and with projectors onto vertex-term-satisfying configurations. It can be checked that all the terms in the Hamiltonian commute with one another.

What is the excitation structure of this semionic version of X-cube model? Fractons can be created exactly as in the (trivalent) toric code version of the X-cube model and have the same properties. The 1D particles (pairs of vertex excitations) can no longer be created simply with an ${\cal X}$ string as in Fig.~\ref{triv_ex}b, as this operator has nontrivial commutation with various plaquette and cube terms along the path of the string. Instead, to create a 1D particle, we need to combine two semion string operators from two intersecting planes along the intersection line. That is we need to supplement the ${\cal X}$ string operator with some extra phase factors depending on the ${\cal Z}$ configuration along the string.

As a result of the structure of the string, the 1D particles have nontrivial ``braiding'' statistics with each other. For example, consider a 1D particle moving in the $x$ direction, which is a composite of a semion in the $xy$ plane and a semion in the $zx$ plane. Suppose that the line on which this 1D particle moves intersects that of another 1D particle moving in the $y$ direction, which is a composite of a semion in $yz$ plane and a semion in the $xy$ plane. The string operators of the two 1D particles anti-commute with each other, because the string operator of one semion on the $xy$ plane anti-commutes with that of another semion on the same plane, while string operators of semions on different planes commute with each other. This anti-commutation of string operators is related to the fact that two such 1D particles can undergo a full braid, so their mutual statistics is well defined. In this case, the two 1D particles have mutual statistics $\theta = \pi$, which contrasts with trivial mutual statistics $\theta = 0$ in the original X-cube model.

In making the above statements, we need to account for the fact that there are different types of 1D particles moving along a given line. For instance, given a 1D particle moving in the $x$ direction, we can attach to it a 2D particle in an $xy$ plane, and obtain a new 1D particle moving in the $x$ direction. Such attachments can change the statistics, so some care is needed to be sure the statistics we find in the semionic X-cube model is really different from that in the original X-cube model.

This can be addressed by demanding that the 1D particles satisfy a certain fusion condition. In particular, 
we require that three 1D particles moving in $x$, $y$, and $z$ directions fuse to a trivial excitation when they meet at a point. This amounts to making a certain natural choice of 1D particle excitations. When this fusion condition is satisfied, 1D particles moving in orthogonal directions indeed have $\theta = 0$ mutual statistics in the original X-cube model, while they have $\theta = \pi$ statistics in the semionic X-cube model.

\section{Discussion} \label{disc}

In this work, we showed how several different fracton topological orders could be realized through coupled layer constructions, both by forming coupled stacks of conventional 2d topologically ordered phases, and by ``stacking'' and coupling a finite number of 3d models with fracton topological order. This perspective allowed us to shed light on the physics of the X-cube model of Ref.\onlinecite{vijay2016fracton}, which we demonstrated could be obtained from a three-dimensional stack of coupled toric codes in the strong coupling limit. These results can be understood through mechanisms we have dubbed ``p-string condensation'' and ``p-membrane condensation,'' in which either one-dimensional strings of particles or two-dimensional membranes of particles are driven to condense. Using these ideas, we constructed two new models of fracton topological order: a semionic generalization of the X-cube model and a phase obtained by inducing p-membrane condensation in a system of four interpenetrating X-cube models, which we dubbed the ``Four Color Cube model.''

We now turn to a discussion of questions raised by our work and potential further avenues of study. A natural first question to ask is whether other known examples of fracton topological order, like the checkerboard model of Ref. \onlinecite{vijay2016fracton} or the fractal topological order of Haah's code \onlinecite{haah2011local}, admit a coupled-layer description similar to the ones presented here. More ambitiously, we can consider asking whether or not {\it all} fracton topological phases can be realized by a coupled layer construction, and if not, whether it is possible to complete a classification of the ones that do. Making progress in this direction would be useful for determining the extent to which fracton topological phases lie ``beyond'' quantum field theory, and may help us to construct more interesting examples of such phases. Along these lines, if possible it would also be interesting to obtain a continuum-picture field-theoretic understanding of p-string and p-membrane condensation, which could further elucidate the degree of the relationship between fracton topological phases and more familiar topological quantum field theories. 

Another natural direction for future work is to perform our p-string condensation procedure for stacks of two-dimensional topological phases other than the toric code and doubled semion examples considered in this paper. We expect the extension to more general types of Abelian topological orders realized in commuting projector models to be fairly straightforward, and can likely be done by following the framework developed in Sec.~\ref{SXC}. Performing our analysis for coupled layers of non-Abelian topological phases may be less straightforward, and at this stage it is unclear exactly what qualitatively new features we might expect to occur in these more general settings.  

Our ability to understand certain fracton topological phases from within the general framework of the theory of two-dimensional topological phases raises the possibility of easily studying other properties of fracton phases, like how their classification is enriched by the presence of symmetries, how symmetry fractionalization and anomalies are classified in fracton phases, and how such phases' edge theories can be constructed. As there already exist a large number of theoretical techniques for studying these properties in conventional topological phases, one could anticipate using our coupled-layer approach to reduce the study of these questions in fracton phases to problems involving conventional topological phases, which could then be solved using existing methods. 

Finally, this paper has taken a rather phenomenological view towards the study of the fusion and braiding properties of the excitations in fracton topological phases, addressing the fusion and statistical properties of each model on a case-by-case basis. It would be useful to develop the fusion and braiding theory of fracton phases on a more general level and to see how such a picture fits in with the commutative algebra methods developed in Ref.\onlinecite{vijay2016fracton}; we plan to pursue this in future work.

\acknowledgments{X.C. would like to thank A. Kubica and J. Mozgunov for discussion. H.M would like to thank Y.-M Lu and Z. Bi for insightful discussion. M.H. and H.M. were supported by the U.S. Department of Energy, Office of Science, Basic Energy Sciences (BES) under Award number DE-SC0014415. X.C. is supported by the Caltech Institute for Quantum Information and Matter and the Walter Burke Institute for Theoretical
Physics. E.L. was supported by the NSF grant PHY-1560023. Some of this work was carried out at the Kavli Institute for Theoretical Physics, which is supported by the National Science Foundation under Grant No. NSF PHY11-25915.}

\appendix

\section{X-cube Hamiltonian from degenerate perturbation theory}
\label{app:perturbation-theory}

Here we apply Brillouin-Wigner perturbation theory to derive the X-cube model in the limit of strongly coupled toric code layers. As briefly discussed at the end of this appendix, essentially same analysis holds in the strong coupling limit of the coupled X-cube model, where we obtain the FCC model. The perturbation theory used to obtain the semionic X-cube model has the same general structure, but some statements in the analysis here assume that the Hamiltonian is a sum of terms that square to unity, which is not true in that case.

We write the Hamiltonian as $H = H_0 + H_1$, where $H_0$ is the $J_z$ coupling term. $H_1 = H_{1v} + H_{1p}$ is the Hamiltonian of decoupled toric code layers, where $H_{1v}$ is the sum of all toric code vertex terms, and $H_{1p}$ the sum of all plaquette terms.   Let $| \psi \rangle$ be an energy eigenstate with energy $E$ that lies in the ground state space of $H_0$ if the perturbation is turned off. We write $| \psi \rangle = | \psi_0 \rangle + | \psi_1 \rangle$, where $| \psi_0 \rangle$ is chosen to be normalized and lies in the ground state space, and $| \psi_1 \rangle$ lies in the orthogonal complement of the ground state space.  We let ${\cal P}$ project onto the ground state space, and $(1 - {\cal P})$ projects onto the orthogonal complement.  Note that ${\cal P}$ commutes with $H_0$.

The Schr\"{o}dinger equation can be written in the form
\begin{equation}
| \psi \rangle = | \psi _0 \rangle + (E - H_0)^{-1} (1 - {\cal P} ) H_1 | \psi \rangle \text{,}
\end{equation}
and iterated to find the formal solution
\begin{equation}
| \psi \rangle =  \sum_{n = 0}^{\infty} \Big[ (E - H_0)^{-1} (1 - {\cal P} ) H_1 \Big]^n | \psi_0 \rangle  \text{.} \label{eqn:bw-wavefn}
\end{equation}
This is not a closed-form solution because $E$ is the unperturbed energy. This does not matter at leading order, and it is common to stop at leading order in Brillouin-Wigner perturbation theory, but we need to go beyond leading order here. Fortunately, the commuting projector nature of $H_1$ will allow us to simplify the perturbation series.

Acting with ${\cal P} H$ on both sides of Eq.~(\ref{eqn:bw-wavefn}), we have
\begin{eqnarray}
{\cal P} H | \psi \rangle &=& E_0 | \psi_0 \rangle \\
&+& {\cal P} H_1  \sum_{n = 0}^{\infty} \Big[ (E - H_0)^{-1} (1 - {\cal P} ) H_1 \Big]^n {\cal P} |  \psi_0 \rangle \text{,} \nonumber
\end{eqnarray}
where we inserted a factor of ${\cal P}$ in front of the $|\psi_0\rangle$ on the right.
We also have
\begin{equation}
{\cal P} H | \psi \rangle = E | \psi_0 \rangle \text{.}
\end{equation}
Comparing these two expressions we see that
\begin{equation}
H_{{\rm eff}} | \psi_0 \rangle = (E - E_0) | \psi_0 \rangle \text{,}
\end{equation}
we have defined the effective Hamiltonian $H_{{\rm eff}}$ to be
\begin{eqnarray}
H_{{\rm eff}} &=&   {\cal P} H_1  \sum_{n = 1}^{\infty} \Big[ (E - H_0)^{-1} (1 - {\cal P} ) H_1 \Big]^{n-1} {\cal P}  \\
&\equiv& \sum_{n=1}^{\infty} \tilde{H}^{(n)}_{{\rm eff}} \text{,}
\end{eqnarray}
where $\tilde{H}^{(n)}_{{\rm eff}}$ is the $n$th term in the series.

This expression for the effective Hamiltonian depends on the energy $E$ of the eigenstate. In particular, this means that $\tilde{H}^{(n)}_{{\rm eff}}$ is not purely of $n$th order in the perturbation. This is an undesirable property that we shall eliminate perturbatively, expanding in corrections to $E$ to get an ordinary Hamiltonian that does not depend on the energy. We will put the effective Hamiltonian in the form
\begin{equation}
H_{{\rm eff}} = \sum_{n = 1}^{\infty} H^{(n)}_{{\rm eff}} \text{,}
\end{equation}
where the tilde has been dropped to signify that $H^{(n)}_{{\rm eff}}$ is truly of $n$th order in the perturbation and does not depend on $E$.

The leading-order contribution is
\begin{equation}
\tilde{H}^{(1)}_{{\rm eff}} = {\cal P} H_{1 v} {\cal P} = H^{(1)}_{{\rm eff}} \text{,}
\end{equation}
because $H_{1p}$ has vanishing projection onto the ground state space.  This term is of first order in the perturbation and does not depend on $E$, so it happens that $\tilde{H}^{(1)}_{{\rm eff}} = H^{(1)}_{{\rm eff}}$.

The behavior at second order is more generic, in that $\tilde{H}^{(2)}_{{\rm eff}} \neq H^{(2)}_{{\rm eff}}$.  
We have
\begin{equation}
\tilde{H}^{(2)}_{{\rm eff}} = {\cal P} H_{1p} \frac{1 - {\cal P} }{E - H_0} H_{1p} {\cal P} .
\end{equation}
In the above expression there is no $H_{1v}$ to the right of the $(1-{\cal P})$ projector, since $H_{1v}$ cannot take states in the ground state space out of the ground state space and since $(1-{\cal P})$ annihilates states in the ground state space. Similarly, there is no $H_{1v}$ to the left of the $(1-{\cal P})$ term. To make the notation more compact we define
\begin{equation}
{\cal D} \equiv \frac{1 - {\cal P}}{E-H_0} \text{,}
\end{equation}
so that $\tilde{H}^{(2)}_{{\rm eff}} = {\cal P}H_{1p} {\cal D}H_{1p}{\cal P}$. It is important to keep in mind that ${\cal D}$ depends on $E$.

To eliminate the $E$-dependence in $\tilde{H}^{(2)}_{{\rm eff}}$, we write $E = E_0 + E^{(1)}$, where $E^{(1)}$ is the first-order correction to the energy. We will not need to include higher-order corrections, because we will see that all such corrections vanish up through fifth order, except for constant corrections that we drop. Now suppose $\tilde{H}^{(2)}_{{\rm eff}}$ acts on an eigenstate of $H^{(1)}_{{\rm eff}}$.  We expand in powers of $ E^{(1)}$, and bring all factors of $E^{(1)}$ to the right of the expression, but still inside the rightmost projector ${\cal P}$. Each factor of $E^{(1)}$ can then be replaced by $H_{1v}$, as a consequence of the form of $H^{(1)}_{{\rm eff}}$ and the fact that we are acting on an eigenstate of $H^{(1)}_{{\rm eff}}$. We can then return $H_{1v}$ to the position where $E^{(1)}$ originally appeared in the expression, because, using the commuting projector structure of $H_1$, $H_{1v}$ commutes with $H_0$, ${\cal P}$ and $H_{1p}$. Since this holds for any eigenstate of $H^{(1)}_{{\rm eff}}$, it holds for all states, and we can expand ${\cal D}$ by
\begin{equation}
{\cal D} = \left( \sum_{n=0}^\infty (-1)^n({\cal D}_0 H_{1v})^n\right){\cal D}_0 \text{,}
\end{equation}
where we define ${\cal D}_0 = (1-{\cal P})/(E_0-H_0)$. Up to this point, we have only used the  properties that $H_1$ is a sum of commuting terms, and that $H_{1v}$ commutes with $H_0$; we note that these properties also hold in the coupled-layer construction of the semionic X-cube model.

Applying this to our expression for $\tilde{H}^{(2)}_{{\rm eff}}$ we obtain
\begin{eqnarray}
\tilde{H}^{(2)}_{{\rm eff}} &=& {\cal P} H_{1p} {\cal D}_0 H_{1p} {\cal P} 
- {\cal P} H_{1p} {\cal D}_0 H_{1v} {\cal D}_0  H_{1p} {\cal P} \\
&+& \sum_{n=2}^{4} (-1)^n {\cal P} H_{1p} \Big( {\cal D}_0 H_{1v} \Big)^n {\cal D}_0 H_{1p} {\cal P} + \cdots \text{.} \nonumber
\end{eqnarray}
The second order contribution of original Hamiltonian to the effective Hamiltonian, ${\cal P} H_{1p} {\cal D}_0 H_{1p} {\cal P} $, is easily seen to be a constant. We therefore drop it and put $H^{(2)}_{{\rm eff}} = 0$. This is convenient because then we then have $E^{(2)} = 0$, which simplifies going to higher orders. The higher order contributions will contribute to $H^{(3)}_{{\rm eff}}$, and so on.

At third order we have
\begin{equation}
\tilde{H}^{(3)}_{{\rm eff}} = {\cal P} H_{1p} {\cal D} H_{1v} {\cal D}  H_{1p} {\cal P} \text{,}
\end{equation}
where only this arrangement of $H_{1v}$ and $H_{1p}$ contributes. Expanding to sixth order, we obtain
\begin{equation}
\tilde{H}^{(3)}_{{\rm eff}} = {\cal P}H_{1p} \left( \sum_{n=1}^4 n(-1)^{n+1} ({\cal D}_0 H_{1v})^n\right) {\cal D}_0H_{1p}{\cal P}
\text{.}
\end{equation}
To determine $H^{(3)}_{{\rm eff}}$, we add the $n=1$ term here with the same order of $H$ contributed by $\tilde{H}^{(2)}_{{\rm eff}}$.  Then we get $H^{(3)}_{{\rm eff}} = 0$; similar cancelations will occur at higher order.

\begin{widetext}

We now give expressions for $\tilde{H}^{(n)}_{{\rm eff}}$ for $n = 4,5,6$, expanding each up through sixth order. The fourth-order term is
\begin{eqnarray}
\tilde{H}^{(4)}_{{\rm eff}} &=& {\cal P} \Big( H_{1p} {\cal D} \Big)^3 H_{1p} {\cal P}
+  {\cal P} H_{1p} ({\cal D} H_{1v})^2  {\cal D} H_{1p}
{\cal P}  \nonumber \\
&\equiv& \tilde{H}^{(4),a}_{{\rm eff}} + \tilde{H}^{(4),b}_{{\rm eff}}  \text{.}
\end{eqnarray}
Expanding $\tilde{H}^{(4),a}_{{\rm eff}} $ up through 6th order we have
\begin{equation} \label{eq:H4aeff3}
\tilde{H}^{(4),a}_{{\rm eff}} =  {\cal P} \Big( H_{1p} {\cal D}_0 \Big)^3 H_{1p} {\cal P} 
+ {\cal P}H_{1p}\left(\sum_{n=1}^2 (-1)^n\sum_{\rm perm} {\rm perm}([{\cal D}_0H_{1p}]^2,[{\cal D}_0H_{1v}]^n)\right) {\cal D}_0 H_{1p}{\cal P},
\end{equation}
where ${\rm perm}([A]^k,[B]^\ell)$ is a term of the form $A^{n_1} B^{m_1} A^{n_2} B^{n_2} \cdots$, with $n_i, m_i$  non-negative integers satisfying $\sum_i n_i = k$ and $\sum_i m_i = \ell$, and the sum runs over all such distinct terms.  For example,
\begin{equation}
\sum_{{\rm perm}} {\rm perm}( [A]^2, [B]^2 ) = A^2 B^2 + A B^2 A + B^2 A^2 + A B A B + B A^2 B + B A B A \text{.}
\end{equation}
In general, ${\rm perm}([A]^k,[B]^\ell)$ runs over $(k+\ell)! / (k!\ell!)$ different terms.

The lowest order term in $H_{1p}$ in the \eqref{eq:H4aeff3} is a constant that we drop. The attentive reader will notice that this term contains a super-extensive contribution to the energy, which we expect would have been canceled had we kept the extensive constant contribution at second order.

For $\tilde{H}^{(4),b}_{\rm eff}$, we have
\begin{equation}
\tilde{H}^{(4),b}_{{\rm eff}} =  {\cal P}H_{1p}\left(({\cal D}_0H_{1v})^2 - 3({\cal D}_0H_{1v})^3 + 6({\cal D}_0H_{1v})^4\right){\cal D}_0H_{1p}{\cal P} \text{.}
\end{equation}

For higher orders, it is helpful to use the more general expression
\begin{eqnarray}
\tilde{H}^{(n)}_{\rm eff} = {\cal P}H_{1p}\left(\sum_{k=0}^{\left \lfloor{n/2-1}\right \rfloor} \sum_{\rm perm} {\rm perm}\left([{\cal D}H_{1p}]^{2k},[{\cal D}H_{1v}]^{n-2k-2}\right)\right){\cal D}H_{1p}{\cal P}.
\end{eqnarray} 
which holds for $n>1$.
Setting $n=5$ for fifth order, we obtain
\begin{eqnarray}
\tilde{H}^{(5)}_{{\rm eff}} &=&  {\cal P}H_{1p} \left(\sum_{\rm perm}{\rm perm}([{\cal D}H_{1p}]^2,[{\cal D}H_{1v}]^1)\right)  {\cal D} H_{1p}{\cal P} + {\cal P} H_{1p} \Big({\cal D} H_{1v}\Big)^3  {\cal D} H_{1p} {\cal P} \\
&=& \tilde{H}^{(5),a}_{{\rm eff}}+\tilde{H}^{(5),b}_{{\rm eff}}. \\
\end{eqnarray}
Expanding each term up to sixth order gives us
\begin{eqnarray}
\tilde{H}^{(5),a}_{{\rm eff}} &=&{\cal P}H_{1p} \left(\sum_{\rm perm}{\rm perm}([{\cal D}_0H_{1p}]^2,[{\cal D}_0H_{1v}]^1) - 2\sum_{\rm perm}{\rm perm}([{\cal D}_0H_{1p}]^2,[{\cal D}_0H_{1v}]^2)\right)  {\cal D}_0 H_{1p}{\cal P}, 
\\
\tilde{H}^{(5),b}_{{\rm eff}} &=&  {\cal P} H_{1p} \Big( ({\cal D}_0 H_{1v})^3  - 4 ({\cal D}_0 H_{1v})^4 \Big){\cal D}_0 H_{1p} {\cal P}.
\end{eqnarray}

Finally, we consider the sixth order term:
\begin{eqnarray}
\tilde{H}^{(6)}_{{\rm eff}} &=&  {\cal P}H_{1p} \left( \sum_{\rm perm}{\rm perm}([{\cal D}H_{1p}]^2,[{\cal D}H_{1v}]^2)\right){\cal D}H_{1p}{\cal P} +{\cal P} H_{1p} \Big({\cal D} H_{1v} \Big)^4 {\cal D} H_{1p} {\cal P} + {\cal P} \Big( H_{1p}{\cal D} \Big)^5 H_{1p} {\cal P} \\
&=& \tilde{H}^{(6),a}_{{\rm eff}}+\tilde{H}^{(6),b}_{{\rm eff}}+\tilde{H}^{(6),c}_{{\rm eff}}.
\end{eqnarray}
Since we will not go beyond sixth order, we can replace all the ${\cal D}$'s with ${\cal D}_0$'s, and write
\begin{eqnarray}
\tilde{H}^{(6),a}_{{\rm eff}} &=& {\cal P}H_{1p} \left( \sum_{\rm perm}{\rm perm}([{\cal D}_0H_{1p}]^2,[{\cal D}_0H_{1v}]^2)\right){\cal D}_0H_{1p}{\cal P},   \\
\tilde{H}^{(6),b}_{{\rm eff}} &=&{\cal P} H_{1p} \Big({\cal D}_0 H_{1v} \Big)^4 {\cal D}_0 H_{1p} {\cal P}, \\
\tilde{H}^{(6),c}_{{\rm eff}}&=& {\cal P} \Big( H_{1p} {\cal D}_0 \Big)^5 H_{1p} {\cal P}.
\end{eqnarray}

\end{widetext}

To compute $H^{(n)}_{{\rm eff}}$ for $n = 4,5,6$, we simply collect terms. We find $H^{(4)}_{{\rm eff}} = H^{(5)}_{{\rm eff}} = 0$, and
\begin{equation}
H^{(6)}_{{\rm eff}} =  {\cal P} \Big( H_{1p} {\cal D}_0 \Big)^5 H_{1p} {\cal P} \text{.}
\end{equation}
In this term, the only combination of six $H_{1p}$ operators which is not a constant, and which survives the leftmost projection onto the ground state space, is the cube operator ${\cal B}_c$ defined in the main text. There are also constant contributions, which we drop.

Thus, up to sixth order the effective Hamiltonian is given by
\begin{equation}
H_{{\rm eff}} = {\cal P} H_{1 v} {\cal P}+ {\cal P} \Big( H_{1p} {\cal D}_0 \Big)^5 H_{1p} {\cal P} \text{.}
\end{equation}
Ignoring constant contributions, this is identical to $H_{XC}$ in Eq.~(\ref{eqn:xcube}). The ${\cal P} H_{1 v} {\cal P}$ term corresponds to the ${\cal A}^{\mu}_i$ term, and the sixth order of $H_{1p}$ builds up to the ${\cal B}_c$ term. The coefficient of the latter is proportional to $1/J_z^5$, because each denominator ${\cal D}_0$ carries a factor of $J_z^{-1}$.

The analysis here applies essentially without modification to the coupled X-cube model described in Sec.~\ref{CXC}, where we obtain the FCC model in the strong coupling limit. There, the $A_i^{\mu}$ terms play the role of $H_{1p}$, and the $B_c$ terms play the role of $H_{1v}$. The same analysis applies because the lowest-order non-constant term formed by taking a product product of $A^{\mu}_i$ operators also appears at sixth order, where the product is over the faces of a cube as described in Sec.~\ref{CXC}.

\section{Topological order and ground state degeneracy of the FCC model}
\label{app:fcc}

Here, we argue that the FCC model has topological order, in the sense that it has a non-trivial ground state degeneracy on the 3-torus, and that the degenerate ground states cannot be distinguished by local measurements. As a byproduct of this discussion, we compute the ground state degeneracy on a finite $L \times L \times L$ torus, where $L$ is the linear system size, setting the spacing between neighboring vertices of the same color to one.

Our strategy is to first construct a complete set of commuting observables (CSCO), and then to use the properties of the CSCO to argue that the model has topological order. We begin by reviewing how this works for the $d=2$ toric code on a $L \times L$ torus, which contains a total of $2 L^2$ Ising spins. The CSCO consists of stabilizers and logical operators. The stabilizers are the $A_i$ vertex and $B_p$ plaquette operators; there are $L$ operators of each type, but they are not all independent, because $\prod_p B_p = 1$ and $\prod_i A_i = 1$.  Therefore there are a total of $2 L^2 - 2$ independent stabilizers.  The stabilizers appear in the Hamiltonian and have eigenvalue $1$ in a ground state, so it follows that $\log_2 {\rm GSD} = 2$. It also follows that the CSCO also includes two logical operators to be complete and label the ground space. One choice for the logical operators is described in Sec.~\ref{sec:gsd}, where they are taken to be string operators $S_m(\gamma_x)$ and $S_m(\gamma_y)$ threading $m$ particles around two independent cycles of the torus, along paths $\gamma_x$ and $\gamma_y$. Only the topological character (more precisely, the homology class) of $\gamma_x$ and $\gamma_y$ is important, because the exact geometry of the paths can be altered by multiplying the string operators by $A_i$ stabilizers.

To argue that the model has topological order, we need to show that a basis for the ground space $\{ | \psi_i \rangle \}$ ($i = 1,\dots,4$) satisfies the property
\begin{equation}
\langle \psi_i | {\cal O} | \psi_j \rangle = 
c_{\cal O} \delta_{i j} \label{eqn:top-order-prop}
\end{equation}
with $c_{\cal O}$ a constant independent of $i$ and $j$.  This equation holds when ${\cal O}$ is any local operator, or more generally a multi-point correlation function of a finite number of local operators. We take the $| \psi_i \rangle$ to be eigenstates of the CSCO. Without loss of generality, we can take ${\cal O}$ to be a product of $X$ and $Z$ Pauli operators, so that ${\cal O}$ either commutes or anticommutes with all the operators in our complete commuting set.  We should assume that ${\cal O}$ commutes with all the stabilizers, since otherwise $c_{\cal O} = 0$ and the property holds trivially.  Then suppose ${\cal O}$ fails to commute with one of the logical operators. In this case, we can ``move the logical operator over'' by multiplying with stabilizers, to get a new logical operator that commutes with ${\cal O}$, as ${\cal O}$ is assumed to be local.  This is only consistent if ${\cal O}$ anticommutes with some stabilizers, which contradicts our assumptions.  Therefore, ${\cal O}$ commutes with all the logical operators. This shows the right-hand side of Eq.~(\ref{eqn:top-order-prop}) vanishes when $i \neq j$.

To complete the argument, we have to consider Eq.~(\ref{eqn:top-order-prop}) when $i=j$, and show the proportionality constant $c_{\cal O} = \langle \psi_i | {\cal O} | \psi_i \rangle$ is independent of $i$.  To do this, we observe that products of $X$ along closed curves winding around the torus can be used to flip logical operators and connect any ground state to any other ground state. These products are simply string operators that thread $e$ particles around the torus, and can also be ``moved out of the way of ${\cal O}$'' by multiplying with $B_p$ stabilizers. Suppose we call such an operator $\chi$, so that $| \psi_j \rangle = \chi | \psi_i \rangle$ and $\chi {\cal O} = {\cal O} \chi$.  Then we have
\begin{eqnarray}
\langle \psi_i | {\cal O} | \psi_i \rangle &=& \langle \psi_i | {\cal O} \chi \chi | \psi_i \rangle \nonumber \\ &=&   \langle \psi_i |  \chi {\cal O} \chi | \psi_i \rangle  = \langle \psi_j | {\cal O} | \psi_j \rangle \text{,}
\end{eqnarray}
which establishes the desired result.

With this review out of the way, we now describe the application of a similar strategy to the FCC model on a $L \times L \times L$ torus. The analysis turns out to be simpler for $L$ odd, so we focus on that case. As an intermediate step, we first count the independent $X$-stabilizers ($B_c$ cube operators) in the X-cube model. There are $L^3$ such operators, but they are not all independent. If $P$ is some $\{ 100 \}$ lattice plane, then we have in total $3L$ constraints among the stabilizers
\begin{equation}
{\cal C}^X_P \equiv \prod_{c \in P} B_c = 1 \text{.}
\end{equation}
But now not all those constraints are independent. We have relation
\begin{equation}
\prod_{P \in \{ xy \text{ planes} \}} {\cal C}^X_P = \prod_{P \in \{ xz \text{ planes} \}} {\cal C}^X_P
= \prod_{P \in \{ yz \text{ planes} \}} {\cal C}^X_P \text{.}
\end{equation}
Each expression above is thus actually the same constraint, and we find $3L - 2$ independent constraints, for a total of $L^3 - 3L + 2$ independent $X$-stabilizers.

This conclusion, and other similar counting problems below, can be checked numerically for reasonably small values of $L$. This is important because the reasoning employed here is not rigorous; in principle, some dependency could have been missed. Our numerical approach is based on mapping the counting of independent stabilizers to a problem in linear algebra over the two-element field ${\mathbb F}_2$. Any product of $X$ Pauli operators can be thought of as an element of the ${\mathbb F}_2$ vector space $V_X \simeq ({\mathbb F}_2)^{3 L^3}$, where vector addition corresponds to operator multiplication. $X$-stabilizers comprise a subspace $S_X \subset V_X$, and the $B_c$ stabilizers make up a spanning set for $S_X$ with $L^3$ elements. Viewing the spanning set as a $L^3 \times 3 L^3$ matrix, the dimension of $S_X$ is the rank of this matrix, which can be determined via row reduction. We used this method to check the counting of independent $X$-stabilizers in the X-cube model for $L = 2, \dots, 8$. Our results agree with those obtained for arbitrary odd $L$ in Ref.~\onlinecite{vijay2016fracton} by rigorous algebraic methods.

We now count independent $X$-stabilizers in the FCC model. Naively, not taking any constraints into account, there are $L^3$ such operators for each underlying X-cube model, for a total of $4 L^3$. For each color, there are $3 L - 2$ constraints involving that color alone, for a total of $4 ( 3 L - 2)$ constraints. In addition, there are constraints that couple all four colors together. Letting $B_0$ be some $X$-stabilizer, we have a constraint
\begin{equation}
{\cal C}^X = \prod_{n_1, n_2} T(n_1 \ba_1 + n_2 \ba_2) B_0 T(n_1 \ba_1 + n_2 \ba_2)^{-1} = 1 \text{,} \label{eqn:111-constraint}
\end{equation}
where $T(\boldsymbol{R})$ is the unitary operator realizing translation by a Bravais lattice vector $\boldsymbol{R}$. The primitive fcc lattice vectors  $\ba_1, \ba_2, \ba_3$ are defined in Sec.~\ref{CXC}. In general, the translation $T(\boldsymbol{R})$ acts in a ``spin-orbit coupled'' manner where the colors are permuted, as described in Sec.~\ref{CXC}. Here, this implies that ${\cal C}^X$ contains contributions from $B_c$ stabilizers of all four colors. It is not obvious \emph{a priori} that ${\cal C}^X = 1$, but this can be shown by a tedious calculation. The product in Eq.~(\ref{eqn:111-constraint}) is over a $\{ 111 \}$ plane, and by symmetry the same constraint holds for any such plane. Na\"{\i}vely this gives $4 L$ constraints, since there are four orientations of $\{ 111 \}$ planes, and $L$ different planes for a given orientation. The actual number of such constraints is $4L - 4$, because taking a product over all $\{ 111 \}$ planes of the same orientation gives a product over all $X$-stabilizers of all four colors, which is not an independent constraint as it can be obtained by taking products of single-color constraints on $\{ 100 \}$ planes. The total number $N_X$ of independent $X$-stabilizers is thus
\begin{eqnarray}
N_X &=& 4 L^3 - 4 ( 3 L - 2) - (4 L - 4) \\
&=& 4 L^3 - 16 L + 12 \text{.}
\end{eqnarray}
This result has been checked numerically for $L = 2,\dots,7$, including even values of $L$.

By electric-magnetic self duality, the number of independent $Z$-stabilizers is equal to the number of independent $X$-stabilizers, $N_Z = N_X$.  Therefore the total number of independent stabilizers is
\begin{equation}
N_S = N_Z + N_X = 8 L^3 - 32 L + 24 \text{.}
\end{equation}
Taking into account the on-site $X^{w_1}_{\ell} X^{w_2}_{\ell}X^{w_3}_{\ell} = 1$ constraint, 
there are eight Ising spins per simple cubic unit cell, for a total of $8 L^3$ Ising spins. We thus infer
the ground state degeneracy
\begin{equation}
\log_2 {\rm GSD} = 32 L - 24 \text{.}
\end{equation}

The counting of stabilizers implies that we need to find $32 L - 24$ independent logical operators, to complete our CSCO. We choose to work with $X$-logical operators (XLOs), \emph{i.e.} those built from products of $X$ operators. Using the ${\mathbb F}_2$ vector space notation described above, we introduce a subspace $C_X$ satisfying $S_X \subset C_X \subset V_X$, which is defined to contain all products of $X$'s that commute with every $Z$-stabilizer. Then $C_X$ is spanned by the union of all $X$-stabilizers and all XLOs. Two XLOs are considered equivalent if they are related by multiplication of $X$-stabilizers; therefore, independent XLOs are associated with elements of the quotient space $L_X \equiv C_X / S_X$. The number of independent XLOs is then given by
\begin{equation}
N_{{\rm XLO}} = \operatorname{dim} L_X = \operatorname{dim} C_X - \operatorname{dim} S_X \text{.} \label{eqn:nxlo}
\end{equation}
Since we expect $N_{{\rm XLO}} = 32 L - 24$, we can use this to complete the CSCO. We choose a putative spanning set for $C_X$ consisting of all $X$-stabilizers and a conjectured generating set of XLOs. If, using this set and Eq.~(\ref{eqn:nxlo}), we find $N_{{\rm XLO}} = 32 L -24$, then we have found a CSCO that consists of all $X$ and $Z$ stabilizers, and the generating set of XLOs.

\begin{figure}
	\includegraphics[width=0.7\columnwidth]{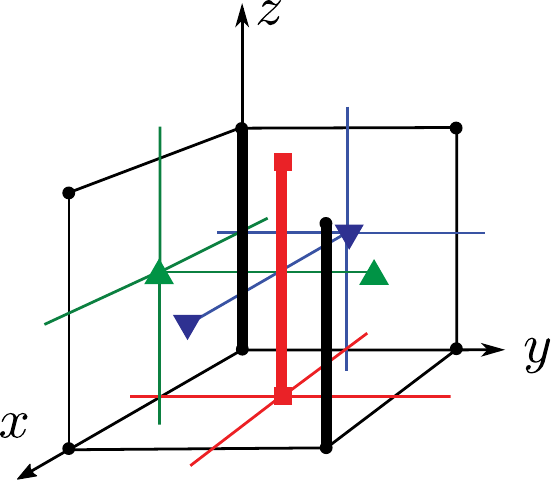}
	\caption{(Color online) Illustration of a type 2 XLO running along the $[110]$ direction. A product is taken over a line of thick-shaded black and red links as shown.  The string extends in the $[110]$ direction, while the links point in the $z$-direction.}
	\label{fig:110}
\end{figure}

We choose a generating set of XLOs that includes two types of operators. Type 1 XLOs are products of $X^w_{\ell}$ of a single color, along a closed straight line (winding once around the torus). We include all 
such XLOs in our generating set. Even though there are $12 L^2$ type 1 XLOs, we find numerically for $L=3,5,7$ that they do not form a complete generating set; that is, when we only include type 1 XLOs in the generating set, we find $\operatorname{dim} C_X - \operatorname{dim} S_X < 32 L - 24$. Evidently, the type 1 XLOs in the generating set are not all independent.

Type 2 XLOs are string operators that run along a $\langle 110 \rangle$ direction; an example is shown in Fig.~\ref{fig:110}.  These strings involve links of two colors that cut transversely to the $\langle 001 \rangle$ plane in which the string lies. We include in the generating set all XLOs running within one $(001)$ plane with arbitrary normal coordinate $z$, and similarly for one $(100)$ and one $(010)$ plane.\footnote{In more detail, for XLOs running in $[110]$ and $[1 \bar{1} 0]$ directions, we consider all red-black XLOs running within one $(001)$ plane with normal coordinate $z$, and all green-blue XLOs running within a neighboring $(001)$ plane with normal coordinate $z + 1/2$.} For $L =3,5,7$, we find that this generating set including both type 1 and type 2 XLOs is enough to generate $32L - 24$ independent XLOs.

To finish arguing for topological order of the FCC model, we have to establish the local indistinguishability of ground states as expressed in Eq.~(\ref{eqn:top-order-prop}). We can follow essentially the same argument for the toric code, noting first that type 1 XLOs can be ``moved away from ${\cal O}$'' by multiplying by $X$-stabilizers. It is not clear how to move type 2 XLOs, but the three planes in which these string operators run can be chosen arbitrarily in such a way that the region on which the local operator ${\cal O}$ is supported is avoided. Finally, electric-magnetic self-duality implies that there is a generating set of $Z$ logical operators (ZLOs) with the same properties as the generating set of XLOs, and these ZLOs can be used to connect different ground states that are eigenstates of the XLOs.

\bibliography{coupled_layer_fracton}

\end{document}